\documentclass[fleqn,usenatbib]{mnras}

\usepackage{newtxtext,newtxmath}
\usepackage[T1]{fontenc}

\DeclareRobustCommand{\VAN}[3]{#2}
\let\VANthebibliography\thebibliography
\def\thebibliography{\DeclareRobustCommand{\VAN}[3]{##3}\VANthebibliography}

\usepackage{graphicx}	% Including figure files
\usepackage{amsmath}	% Advanced maths commands
\usepackage{xcolor}

\newcommand{\todo}[1]{\textcolor{black}{#1}}

\defcitealias{montypaperI}{Paper I}
\title[Precision Abundances in NGC~288 \& NGC~362 II]{Peeking beneath the precision floor - II. Probing the chemo-dynamical histories of the potential globular cluster siblings, NGC~288 and NGC~362}

\author[S. Monty et al.]{Stephanie Monty,$^{1,2,3}$\thanks{E-mail: sm2744@cam.ac.uk}
David Yong,$^{2,3}$,
Davide Massari$^{4}$,
Madeleine McKenzie$^{2,3}$,
GyuChul Myeong$^{5}$,
\newauthor
Sven Buder$^{2,3}$,
Amanda I. Karakas$^{6,3}$,
Ken C. Freeman$^{2,3}$,
Anna F. Marino$^{7,8}$,
Vasily Belokurov$^{1}$,
\newauthor
N. Wyn Evans$^{1}$
\\
$^{1}$ Institute of Astronomy, University of Cambridge, Madingley Rd, Cambridge, CB3 0HA, UK\\
$^{2}$ Research School of Astronomy and Astrophysics, Mt Stromlo Observatory, Weston Creek, ACT 2611, Australia\\
$^{3}$ ARC Centre of Excellence for All Sky Astrophysics in 3 Dimensions (ASTRO 3D), Australia\\
$^{4}$ INAF - Osservatorio di Astrofisica e Scienza dello Spazio di Bologna, Via Gobetti 93/3, I-40129 Bologna, Italy\\
$^{5}$ Harvard-Smithsonian Center for Astrophysics, 60 Garden Street, Cambridge, MA 02138, USA\\
$^{6}$ School of Physics \& Astronomy, Monash University, Clayton, VIC 3800, Australia\\
$^{7}$ Instituto Nazionale di Astrofisica - Osservatorio Astronomico di Padova, Vicolo dell'Osservatorio 5, IT-35122,  Padua, Italy\\
$^{8}$ Instituto Nazionale di Astrofisica - Osservatorio Astrofisico di Arcetri, Largo Enrico Fermi, 5, IT-50125, Firenze, Italy
}

\date{Accepted XXX. Received YYY; in original form ZZZ}

\pubyear{2015}

\begin{document}
\label{firstpage}
\pagerange{\pageref{firstpage}--\pageref{lastpage}}
\maketitle

\begin{abstract}
The assembly history of the Milky Way (MW) is a rapidly evolving subject, with numerous small accretion events and at least one major merger proposed in the MW's history. Accreted alongside these dwarf galaxies are globular clusters (GCs), which act as spatially coherent remnants of these past events. Using high precision differential abundance measurements from our recently published study, we investigate the likelihood that the MW clusters NGC~362 and NGC~288 are galactic siblings, accreted as part of the Gaia-Sausage-Enceladus (GSE) merger. To do this, we compare the two GCs at the 0.01~dex level for 20+ elements for the first time. Strong similarities are found, with the two showing chemical similarity on the same order as those seen between the three LMC GCs, NGC~1786, NGC~2210 and NGC~2257. However, when comparing GC abundances directly to GSE stars, marked differences are observed. NGC~362 shows good agreement with GSE stars in the ratio of Eu to Mg and Si, as well as a clear dominance in the $r$- compared to the $s$-process, while NGC~288 exhibits only a slight $r$-process dominance. When fitting the two GC abundances with a GSE-like galactic chemical evolution model, NGC~362 shows agreement with both the model predictions and GSE abundance ratios (considering Si, Ni, Ba and Eu) at the same metallicity. This is not the case for NGC~288. We propose that the two are either not galactic siblings, or GSE was chemically inhomogeneous enough to birth two similar, but not identical clusters with distinct chemistry relative to constituent stars.
\end{abstract}

%%%%%%%%%%%%%%%%%%%%%%%%%%%%%%%%%%%%%%%%%%%%%%%%%%

%%%%%%%%%%%%%%%%% BODY OF PAPER %%%%%%%%%%%%%%%%%%

\section{Introduction}
\label{sec:intro}
Recent data releases from the Gaia mission \citep[DR2, EDR3, DR3,][]{gaiadr2, gaiaedr3, gaiadr3} have triggered a revolution in our understanding of the assembly history of the Milky Way (MW). Leveraging this unprecedented dataset of nearby stars, a plethora of past merger events between the MW and dwarf galaxies (dGals) have been proposed \citep[e.g.][]{belokurov18, haywood18, helmi18, myeong19, massari19, kraken, koala, heracles, thamnos, arjuna}. Along with the dGals themselves, each of these merger events is likely to have brought in a system of associated globular clusters (GCs) \citep{searlezinn, mackey04, abadi06, law10, trujillo21, shirazi22}. And while these accreted dGals have long since been disrupted, distributing their constituent stars across the MW, some members of their GC systems have likely remained intact. Identifying and characterising the GC systems of these long-gone dGals could provide a chemo-dynamical link to these past merger events.

Prior to Gaia, classifying the MW's population of GCs as a means of inferring their origins was largely reliant on the GC age-metallicity relation \citep[AMR,][]{amr09, forbes10}. Coupling the AMR with GC kinematics subsequently strengthened the proposed distinction between accreted vs in-situ GCs \citep{leamanamr}. Old, metal-poor clusters were found to generally exhibit halo kinematics and were proposed to have been accreted, while younger, more metal-rich clusters showed disc kinematics and were proposed to have formed in-situ with the MW. The separation of these populations was further supported via their different [Si,Ca/Fe] abundances as shown by \citet{recioblanco18}, with old, metal-poor clusters occupying a plateau in [Si,Ca/Fe].

Thanks to the incredibly high quality of the Gaia DR2/DR3 data \citep{gaiadr2, gaiadr3}, revised GC parallaxes and proper motions are now available to calculate precise cluster orbits. Using revised GC orbital characteristics (e.g. orbital energy, the $z$-component of angular momentum and/or orbital actions), studies by \citet{myeong18}, \citet{myeong19} and \citet{massari19} classified GCs as having been accreted alongside proposed accretion events, including the Sequoia and Gaia-Sausage-Enceladus \citep[Gaia-Sausage/GSE,][]{belokurov18, helmi18} \footnote{Note that while the Gaia Sausage and Enceladus events share many member stars in common and occupy similar, but not identical regions of $E$ vs $L_{z}$ space, they have distinctly different progenitor scenarios and in-fall orbits. We will endeavour to discuss the two separately where possible.}, or as having formed in-situ alongside the progenitor of the MW disc. This has since expanded with the discovery of additional accretion events and the release of EDR3 \citep{callingham22, malhan22a, limberg22}.

Thus far, dynamical associations of GCs to merger events has involved integrating GC orbits within a static potential (without a live halo composed of particles). The associations are then made when clusters are found to group together in the dynamical spaces associated with proposed progenitors. However, groups in dynamical spaces sometimes overlap, making the associations uncertain. Moreover, further complexity can be introduced when considering live halos, particularly in energy vs. $z$-angular momentum space \citep{pagnini22}. Unfortunately, until now their chemical abundances have not yielded greater clarity, revealing no distinct differences between clusters from proposed events \citep{hortagcs}. 

Using catalogues of GC stars within the APOGEE survey \citep{apogee}, \cite{hortagcs} tagged GCs to different accretion events in energy vs. $z$-angular momentum space following the convention of \cite{massari19}. When examining the average cluster abundances in the $\alpha$-elements Si and Mg (elements largely associated with nucleosynthsis via core-collapse supernovae), they found no distinctions between GCs tagged to different events. This is despite a clear difference in $\alpha$-element abundances being seen between in-situ and accreted MW GCs \citep{recioblanco18} and is likely the result of limited chemical abundance precision. Given that studies have shown that GSE field stars occupy both a unique region of chemical as well as dynamical space with respect to MW field stars  \citep{haywood18, matsuno19, monty20, feuillet20, matsuno21, aguado21, feuillet21, buder22, horta22}, and evidence of the chemical distinction among proposed accreted substructure is emerging \citep{horta22, matsuno22a, matsuno22b}, chemical homogeneity of different GC populations seems unlikely.

In addition to the wealth of quality information large spectroscopic surveys can provide, we owe a major discovery in the MW to a small study that pushed the chemical precision floor to a new limit. The discovery by \citet{nissen10}  that the MW hosts two distinct halo populations, a high- and low-$\alpha$ sequence (and the assertion that these represent the in-situ and accreted halo populations) was only possible due to measurement errors on the order 0.01~dex. This was achieved through performing differential abundance measurements, a technique pioneered in the study of solar twins \citep{melendez09}. Another example is the discovery that the seemingly disc-like GC NGC~6752 hosts star-to-star abundance variations at the two sigma level for every element measured, in addition to significant correlations between unexpected elements \citep{yong2013}. This can only be explained by invoking a specific enrichment pattern in the (pre)proto-cluster environment, providing a direct link to the cluster’s parent galaxy.

Continuing with the exploration of MW GCs in the high precision regime, we conducted a differential chemical abundance study of the two clusters NGC~288 and NGC~362. The two clusters are particularly interesting as they have nearly identical metallicities \citep[NGC~288: $\mathrm{[Fe/H]}\sim-1.39$,  NGC~362: $\mathrm{[Fe/H]}\sim-1.33$,][]{shetrone00} and orbital characteristics \citep[e.g. NGC~288: ($r_{\mathrm{peri}}$, $r_{\mathrm{apo}}$) = (3.33, 13.01)~kpc, NGC~362: ($r_{\mathrm{peri}}$, $r_{\mathrm{apo}}$) = (1.05, 12.48)~kpc and both are on weakly retrograde orbits,][]{baumgardtorb} and yet they are chemically distinct \citep{shetrone00, carretta13} and show clear signs of different dynamical evolution \citep{dalessandro13, piatti18, sollima22}. Using integrals of motion space \citep{helmi00}, \citet{massari19} classified both clusters as having been accreted in the GSE merger, while \citet{myeong18, myeong19} lists NGC~362 as having a \textit{possible} association with the Gaia-Sausage event and NGC~288 as having no association.

This is the second paper in a series using differential analysis to probe the chemistry of NGC~288 and NGC~362. In our first paper \citep[][hereafter Paper I]{montypaperI}, we explored the chemistry in the two clusters in the context of GC chemical evolution. Our analysis revealed statistically significant dispersion in several elements in both clusters including, \ion{Fe}{I} and heavy elements from the slow neutron capture ($s$-process) nucleosynthetic group. 

In addition to a spread in the $s$-process elements, NGC~362 displayed a spread in the heavy element Eu, a member of the rapid neutron capture ($r$-process) nucleosynthetic group \citep{burbidge}. Furthermore, using a spread in $s$-process elements, we identified two distinct groups within NGC~362 which we believe are associated with separate star formation (SF) events. The older, $s$-process weak population was found to be dominated by the $r$-process with a differential $\Delta^{\mathrm{La}}-\Delta^{\mathrm{Eu}}$\footnote{Note that we use the $\Delta^{\mathrm{X}}$ notation when referring to \textit{differential} or relative abundance ratios. This is analogous to the traditional [X/Fe] notation used in spectroscopic studies. Further explanation is given in Sec.~\ref{sec:paperI}} ratio of $-0.16\pm0.06$. We then proposed that the $r$-process enrichment and dispersion found within the older group were primordial. This discovery motivated further research into whether primordial GC abundances reflect the abundances of their host galaxies. Can the two be linked?

In this study, we explore the possibility that the two GCs, NGC~288 and NGC~362, are galactic siblings accreted as part of the same event. To investigate this question, we make use of high precision differential abundance measurements from our previous study of NGC~288 and NGC~362 (\citetalias{montypaperI}, discussed in Sec.~\ref{sec:paperI}) to compare and contrast the two. Sec.~\ref{sec:chemodyn} explores both scenarios using chemo-dynamical evidence that suggests that the GCs \textit{are} (Sec.~\ref{sec:dynsib} and \ref{sec:chemsupport}) and \textit{are not} siblings (Sec.s~\ref{sec:dynagainst} and \ref{sec:chemevidagainst}). Without assuming an accretion scenario, in Sec.~\ref{sec:gce} we assess the fit of a GSE-tailored galactic chemical evolution model to six different abundance ratios in our GCs. Finally, in Sec.~\ref{sec:sum} we summarise the results of our investigation and provide conclusions as to the potential origin of the two GCs in Sec.~\ref{sec:conc}.

\section{Summary of Observations and Analysis from Paper I}\
\label{sec:paperI}
In \citetalias{montypaperI} we used the technique of differential abundance analysis to measure Na, Al, Mg, Si, Ca, Ti, Cr, Fe, Co, Ni, Cu, Zn, Sr, Y, Zr, Ba, La, Ce, Nd and Eu in NGC~288 and NGC~362. Six \todo{red giant branch (RGB) stars from the tip of the RGB} in NGC~288 and eight from NGC~362 from the study of \citet{shetrone00} were selected to re-observe using the VLT/UVES \citep[$R\sim110,000$,][]{uves}. The star mg9 \citep[also known as B3169:][]{buonanno86} was selected from the pioneering study of \citet{yong2013} to use as a reference star for the analysis. 

Briefly, differential abundance analysis exploits the selection of ``stellar siblings'', stars with similar stellar parameters (effective temperature, $\mathrm{T}_{\mathrm{eff}}$, surface gravity, log~$g$ and metallicity expressed as [Fe/H]) to minimise systematic errors amongst the program and reference stars \citep[see ][for a review of the technique]{nissen18}. The software \texttt{q2} \citep{q2} was used to perform the analysis, in tandem with the radiative transfer code \texttt{MOOG} \citep{moog} using \texttt{MARCS} \citep{marcs} model atmospheres. Ultimately, relative abundance errors on the order 0.01-0.02~dex were recovered in both clusters with the smallest uncertainties being associated with \ion{Fe}{I} in both clusters (0.01~dex in NGC~288 and 0.016~dex in NGC~362.) Further details of the abundance analysis, determination of stellar parameters and line list can be found throughout Sec.~2 in \citetalias{montypaperI}.

\section{Chemodynamics: In the Context of Milky Way Evolution}
\label{sec:chemodyn}
In this section we use chemo-dynamical evidence to explore two possibilities regarding the origin of the clusters. The first, is that the two GCs formed in the \textit{same} dGal and were accreted as part of the \textit{same} merger event, in essence that they are ``galactic siblings''. And the second, is that they did not originate from the same dGal and instead joined the MW via different accretion scenarios. 

\begin{figure}
	\includegraphics[width=\linewidth]{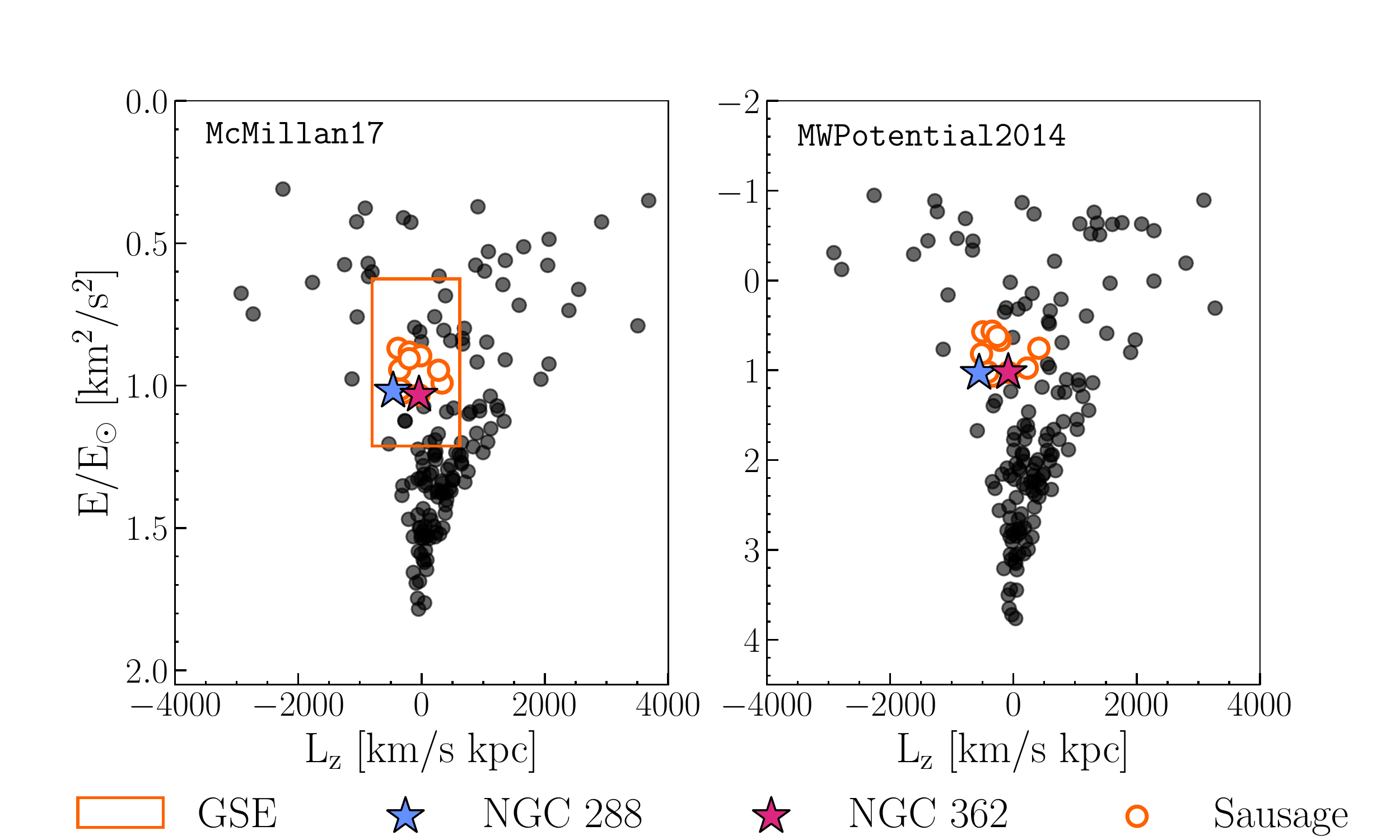}
	\caption{Energy (E in km$^{2}$/s$^{2}$) scaled by solar energy vs. angular momentum (L$_{z}$ in km/s${2}$~kpc) for the majority of MW GCs calculated using the \texttt{McMillan17} MW potential \citep[as in the studies of][]{myeong18, massari19}, integrating for 2~Gyr (\textit{left}). The extent of GSE in E vs. L$_{z}$ as defined by \citet{massari19} is marked in orange, while the Sausage GCs identified by \citet{myeong18} are shown as the open orange circles. NGC~288 and NGC~362 are shown as the blue and pink stars respectively. The same information is shown for GCs integrated in the \texttt{MWPotential2014} MW potential (\textit{right}).}
    \label{fig:elz}
\end{figure} 

\subsection{Accretion Scenario I: Galactic Siblings}
\label{sec:galsib}
We begin our exploration of the cluster accretion histories by discussing the possibility that they formed alongside the same host dGal and therefore joined the MW as part of the same accretion event. As discussed in Sec.~\ref{sec:intro}, this scenario has been proposed for these two clusters by \citet{massari19}, with the two clusters brought in as part of the GSE merger. \citet{callingham22} and \citet{limberg22} further support this classification, assigning both NGC~288 and NGC~362 as GSE GCs with 100\% certainty in their models. \citet{malhan22a} also support the sibling scenario, but assign NGC~288 and NGC~362 to a separate accretion event termed ``Pontus''. Pontus is proposed to have been accreted $\sim1$~Gyr before GSE \citep{malhan22b, hammer23}. To support this theory, we experiment with classifying the clusters dynamically within a four component MW potential and compare the scale of chemical similarities between the two clusters to those found between known galactic siblings within the Large Magellanic Cloud (LMC) GC system.

\begin{table}
	\centering
	\caption{Characteristics of the two Dehnen bars (\texttt{DBP}) and our Ferrers (\texttt{FP}) potential bulge-bar used to explore perturbations to the orbits of our two GCs in Sec.~\ref{sec:dynsib}. The two Dehnen bars are defined by the bar pattern speed ($\Omega_{b}$), bar radius ($R_{b}$) and bar strength ($A_{f}$). Characteristics for the strong Dehnen bar are taken from the simulation parameters of \citet{monaribar} and characteristics for the weak bar from the range of derived values in \citet{wegg15}. Our Ferrers (\texttt{FP}) potential ellipsoid is defined by the total bar+bulge mass M$_{\mathrm{bar+bulge}}$, scale radius ($a$) and major and minor-axis ratio ($e$), with the major axis sitting in the Y-X plane.}
	\label{tab:barchar}
	\begin{tabular}{lllllll} % four columns, alignment for each
		\hline
		Potential & $\Omega_{b}$  & $R_{b}$ & $A_{f}$ & M$_{\mathrm{bar+bulge}}$ & $a$ & $e$ \\
		    & km/s/kpc  & kpc   & km$^{2}$/s$^{2}$ & $10^{10}$~M$_{\odot}$    & kpc   &  \\
		\hline
		\texttt{DBP} (strong)    & 52    & 3.4   & 2101  & ...   & ...   & ... \\
		\texttt{DBP} (weak) & 40   & 5.1   & 621   & ... & ...   & ... \\
		\texttt{FP}    & 40    & 5.1  & ...   & 2   & 5 & 0.2 \\
		\hline
	\end{tabular}
\end{table}

\subsubsection{Dynamical Support for the Sibling Scenario}
\label{sec:dynsib}
The studies of \citet{massari19} and \citet{callingham22} assign cluster association primarily based on the two GCs occupying a similar region of energy ($E$) vs the $z$-component of angular momentum ($L_{z}$) space. In the case of \citet{callingham22} and \citet{limberg22}, actions $\textbf{J}$ were also used. The location of the two GCs in solar-scaled $E$ vs. $L_{z}$ space is shown in Fig.\ref{fig:elz} after integration in the \texttt{McMillan17} potential of \citet{mcmillan17} and the MW potential \texttt{MWPotential2014} of \citet{galpy} for 2~Gyr. The solar energy values used to scale the y-axes in Fig.~\ref{fig:elz} were calculated in the respective potentials, illustrating a fundamental difference between the two - the component masses. A collection of MW GCs is also plotted in the background of Fig.~\ref{fig:elz}. An orange box is placed around GSE using the limits from \citet{massari19} (calculated in the \texttt{McMillan17} potential), while orange open circles mark all the proposed Sausage GCs from \citet{myeong18} - note that there is not complete overlap between the two proposed events (see the footnote in Sec.~\ref{sec:intro}). The similar location of the two GCs in $E$ vs. $L_{z}$ implies that the two must share many orbital characteristics in common, which is indeed the case. 

Using the Gaia EDR3 proper motions and parallax distances for the GCs as determined by \citet{vasilievbaumgardt} we have re-determined the orbital parameters of the two GCs. To do this we use the \texttt{MWPotential2014} potential in \texttt{galpy} \citep{galpy} and \todo{orient ourselves in the usual way. That is, we assume the Sun is 20.8~pc above the galactic plane \citep{bennett19}, the distance to the solar circle is 8.178~kpc \citep{gravity19} and that circular velocity at the solar circle is 229~km/s \citep{eilers19}.} We integrate the orbits forwards and backwards for 5~Gyr and recover similar pericentric and apocentric radii for the two GCs ($r_{\mathrm{peri}}=0.25$~kpc and $r_{\mathrm{apo}}=10.90$~kpc for NGC~288 and $r_{\mathrm{peri}}=0.38$~kpc and $r_{\mathrm{apo}}=13.17$~kpc for NGC~362.) Additionally, the two GCs are found to have similar orbital eccentricities (NGC~288: $e=0.64$ and NGC~362: $e=0.84$) and orbital periods. Finally, in the axisymmetric, time-invariant \texttt{MWPotential2014} potential, the GCs share similar values of $E$, $L_{z}$ and actions as one would expect.

\begin{figure*}
	\includegraphics[scale=0.325]{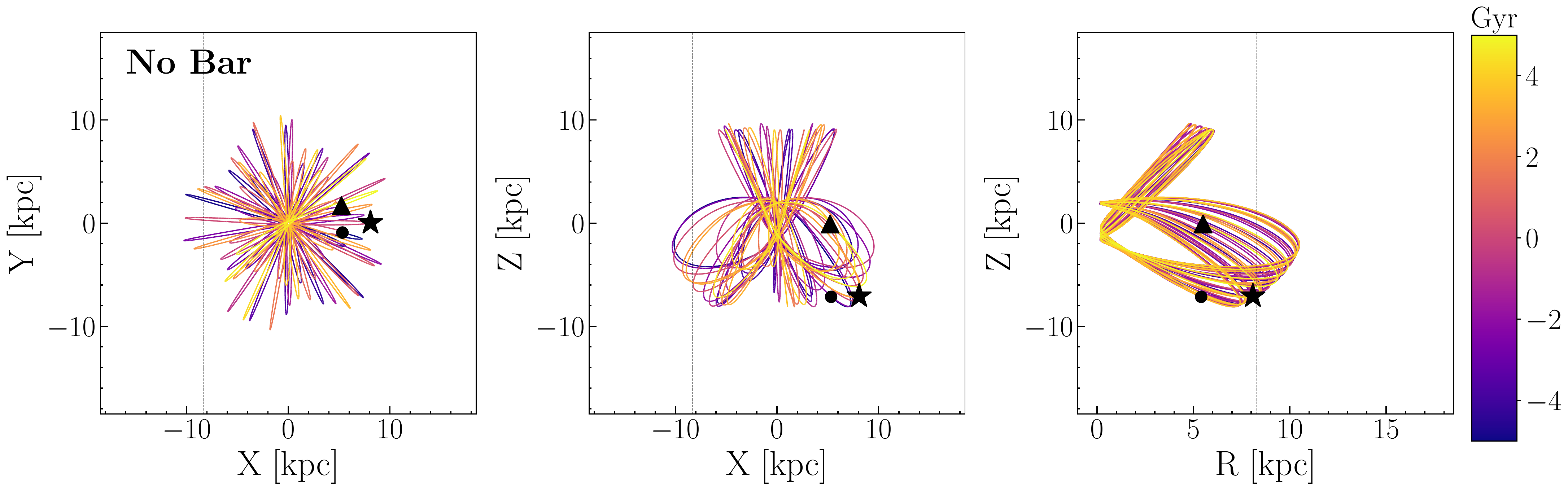}
	\includegraphics[scale=0.325]{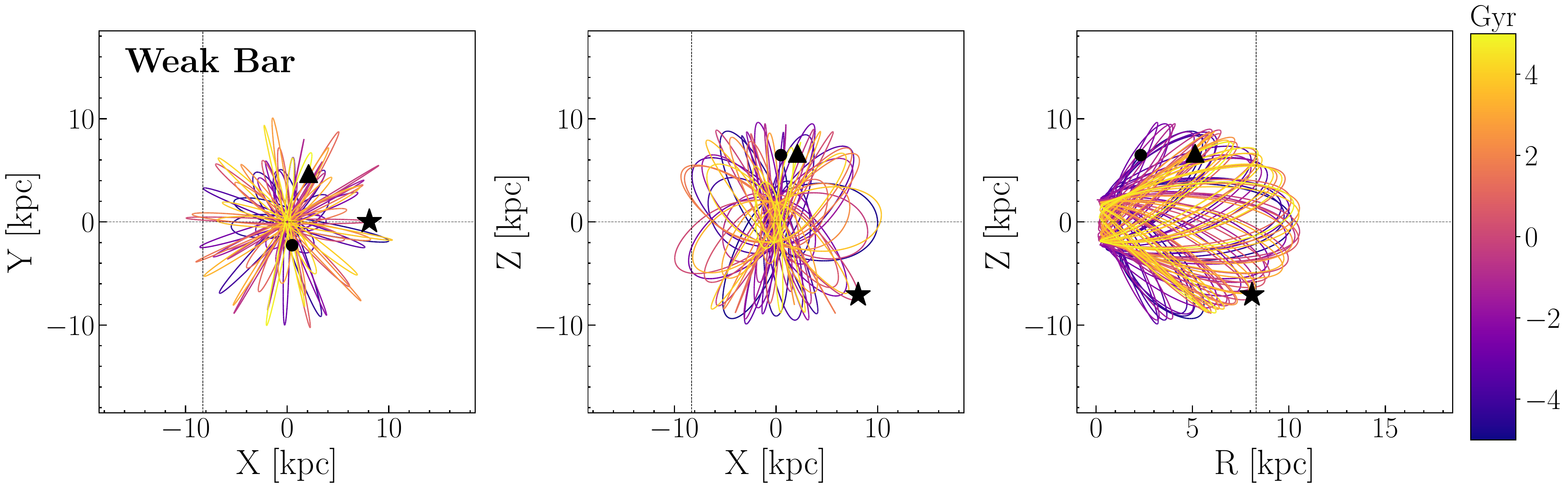}
	\includegraphics[scale=0.323]{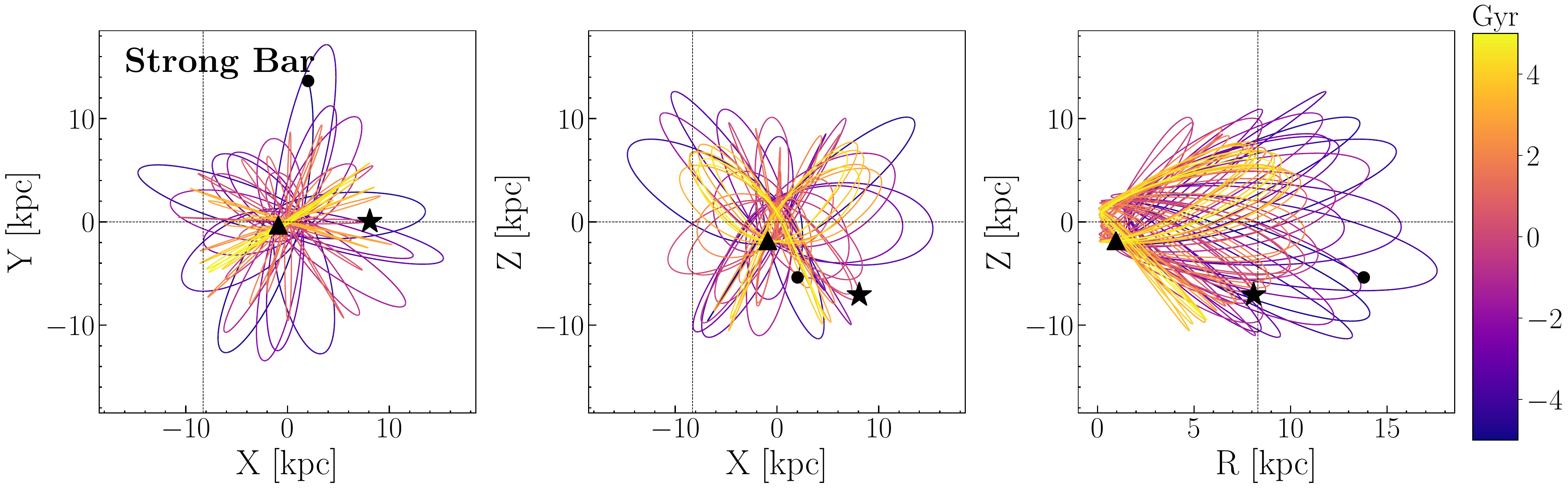}
    \caption{\label{fig:orbitdifs} Example orbits for NGC~288 integrated forwards and backwards for 5~Gyr in three different potentials. The position of the cluster 5~Gyr ago is marked with a circle, the present day position is marked with a star and the position 5~Gyr in the future is marked with a triangle. The trajectory is coloured by time and presented (from left to right) in the Y-X, Z-X and Z-R planes. \textit{Top}, the orbit integrated in \texttt{MWPotential2014}, \textit{middle}, integrated in \texttt{MWPotential2014+DehnenBarPotential} with weak bar characteristics, \textit{bottom} integrated in \texttt{MWPotential2014+DehnenBarPotential} with strong bar characteristics. Details of the bar characteristics are given in Section~\ref{sec:dynsib}. The position of the sun is marked with dashed lines.}
\end{figure*}

Both \citet{massari19} and \citet{callingham22} derive their ``tagging'' characteristics, $E$ and $L_{z}$ (and $\textbf{J}$) under the assumption of a time-invariant, axisymmetric potential (the conditions under which all of these quantities are conserved). Both studies assume the MW potential of \citet{mcmillan17} as implemented in \texttt{AGAMA} \citep{agama}\footnote{The \citet{mcmillan17} potential does not include a component for the bar and instead only models the bulge, disc and dark matter (DM) halo. See Tables 1 and 4 in \citet{mcmillan17} for characteristics of each component.}. In reality, the MW potential is not time invariant, nor symmetric in the inner regions due to the presence of the MW bar.

Many of the orbits of the MW GCs, including NGC~288 and NGC~362 pass within the inner region of the MW where they are subject to the effects of the bar \citep[recent estimates of the bar co-rotation radius are $6.5<R_{\mathrm{CR}}<7.5$~kpc,][] {clarkebar}. This raises the issue of whether the presence of a bar would impact the tagging of the two clusters by changing the orbital characteristics. While testing the impact of a time-variable potential is non-trivial, we can explore the impact of adding a model bar to our MW potential.

We explored two different models for the bar, using the \texttt{galpy} implementation of Dehnen's Bar in 3D \citep[\texttt{DehnenBarPotential},][]{dehnenbar, monaribar} and the \texttt{galpy} implementation of Ferrers potential \citep[\texttt{FerrersPotential,}][]{ferrers}. In the case of the Dehnen potential this is added to the three component potential of \texttt{MWPotential2014} (representing the MW bulge, disc and dark matter halo). We adopt \texttt{MWPotential2014} for this exercise as the form of the potential components make the addition of a static bar fairly simple. In the case of Ferrers potential, which models both the bulge and the bar as an ellipsoid, the potential is built from the halo and disc components of \texttt{MWPotential2014} with the addition of \texttt{FerrersPotential}.

\begin{figure}
    \centering
    \includegraphics[width=\linewidth]{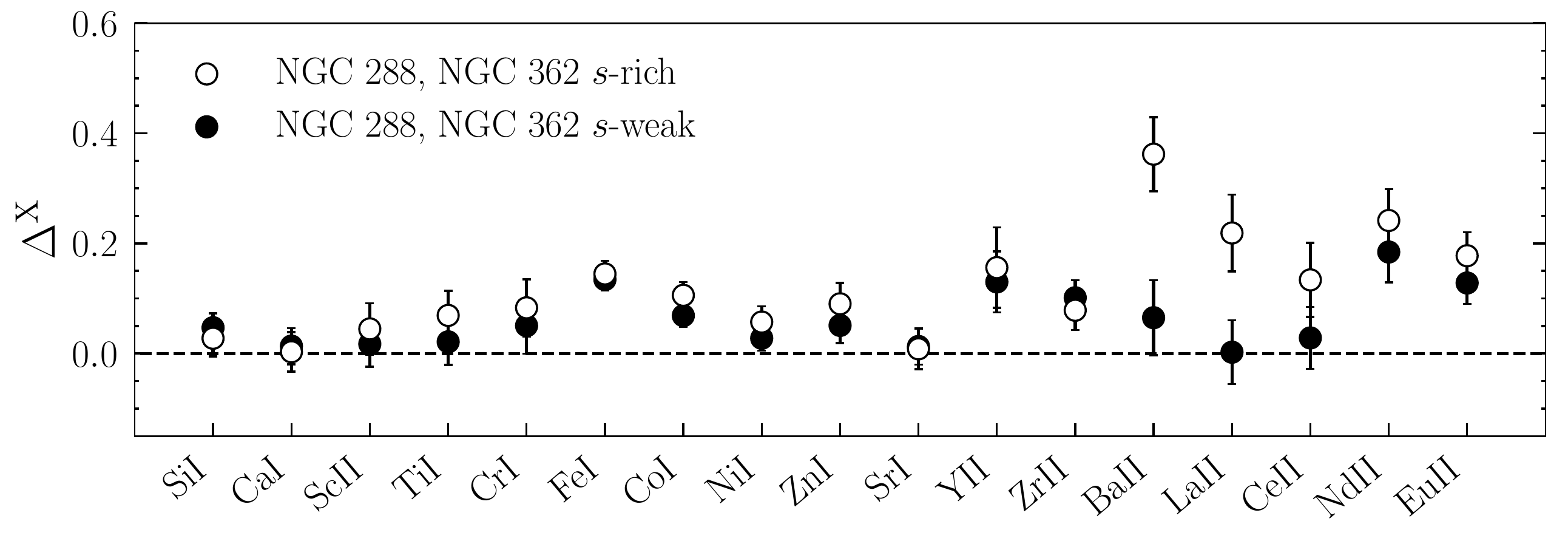}
    \includegraphics[width=\linewidth]{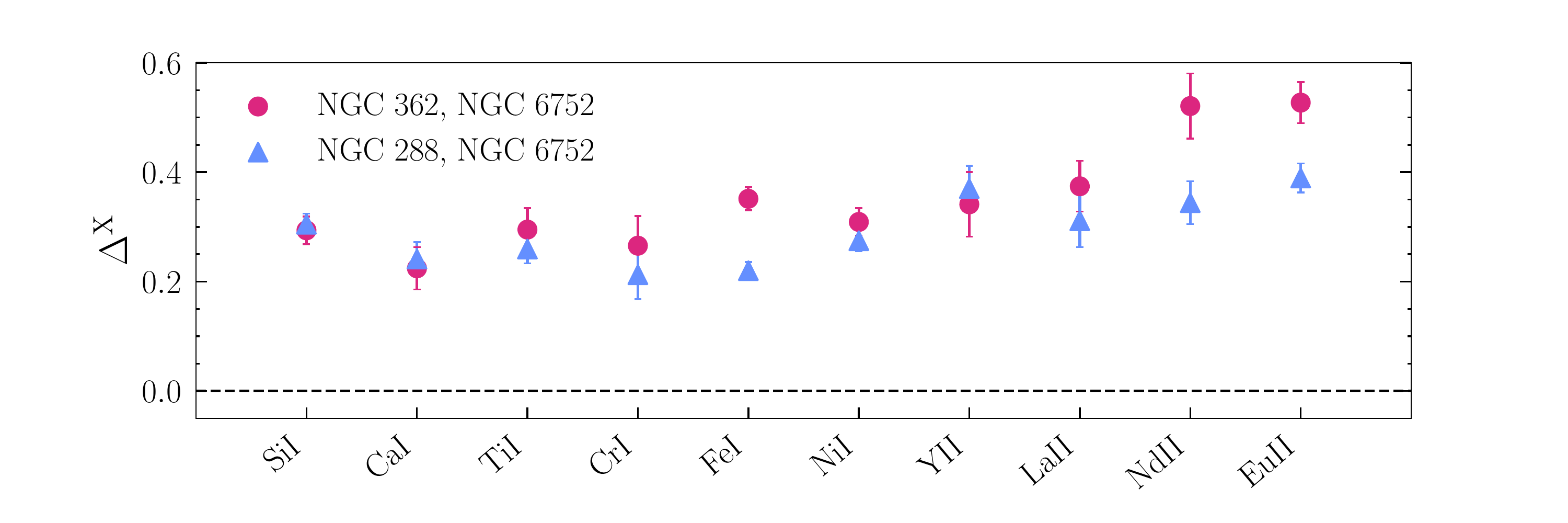}
    \caption{Comparison of the average abundance differences between NGC~288 and the two $s$-process groups in NGC~362 \textit{(top)}. The comparison with the $s$-weak group is shown using the filled circles, while the comparison between NGC~288 and the $s$-rich group is shown using open circles. \textit{Bottom,} the differences between the two GCs in our study and the disc-like GC, NGC~6752 from \citet{yong2013} (when considering all the stars in NGC~362, with the exception of the extremely $s$-process enhanced star NGC362-1441).}
    \label{fig:clusterdif1}
\end{figure}

To build our Dehnen bars we select characteristics for both a strong and weak bar and investigate the impact of both on the cluster parameters. For Ferrers potential, we only experiment with a weak bar. Characteristics of the three bar models are listed in Table~\ref{tab:barchar}. Three example orbits for NGC~288 over the 10~Gyr total integration time are shown in Fig.~\ref{fig:orbitdifs}, in the Y-X, Z-X and Z-R planes for different choices of the potential. NGC~288 was chosen to demonstrate the orbital differences as the changes to the orbit were more severe overall when compared to NGC~362 (for completeness the orbital changes to NGC~362 are shown in Fig.~\ref{fig:ngc362orbs}). The top-most plot shows the orbit integrated in the \texttt{MWPotential2014} potential, the middle and bottom-most plots show the orbit integrated in the \texttt{MWPotential2014+DehnenBarPotential} potential with weak and strong bar characteristics respectively. 

Despite the orbits showing some differences in Fig.~\ref{fig:orbitdifs}, across the three different implementations of the Dehnen bar potential, no significant change in the ``tagging'' parameters $E$ and $L_{z}$ was observed. To explore whether the tagging of the two GCs has changed under the addition of the bar, we focus on the change in $L_{z}$ across potentials. This is because energy is highly sensitive to the mass differences between the \texttt{MWPotential2014} and the \texttt{McMillan17} potential.

Subject to the \texttt{MWPotential2014} potential, the angular momentum of the two GCs is $L_{z}=-41.39$~km/s kpc (NGC~288) and $L_{z}=-89.14$~km/s kpc (NGC~362), well within the range associated with the GSE merger as identified by \citet{massari19} ($-800<L_{z}<620$~km/s kpc). Subject to our bar modeled using Ferrers potential, this changes to $L_{z}=-187.83$~km/s kpc for NGC~288 and $L_{z}=-93.31$~km/s kpc for NGC~362. Finally, under our weak and strong Dehnen Bar respectively, the $L_{z}$ ranges from -31.49~km/s kpc to -70.48~km/s kpc for NGC~288 and -70.79~km/s kpc to -93.06~km/s kpc for NGC~362. Again, these values are well within the range of $L_{z}$ associated with GSE as given by \citet{massari19} and \citet{helmi18}. Conceptually one can see the impact of the bar ``spinning up'' the GCs as they interact with the bar, with the stronger bar causing more spin up. As an experiment, we integrated both clusters forwards for an additional 5~Gyr and saw an additional increase of 30-50~km/s kpc in $L_{z}$ in both clusters.

\todo{\emph{Our first conclusion is that in the presence of a bar as we have modeled it and \textit{without} exploring a time variable potential, the GCs retain their dynamical association to GSE, supporting the idea that they could be galactic siblings.}}

\subsubsection{Chemical Support for the Sibling Scenario}
\label{sec:chemsupport}
To investigate the chemical similarities between the two clusters we split NGC~362 into the two $s$-process groups ($s$-rich and $s$-weak) discovered in \citetalias{montypaperI} and compare both groups to NGC~288. As discussed in Sec.~\ref{sec:intro}, two distinct groups were found within NGC~362 separated by $\sim0.3$~dex in their mean $s$-process abundances ($\Delta^{\mathrm{Y, Ba, La}}$). Each group contains four stars, with the $s$-rich group also containing the extremely $s$-process enhanced star NGC362-1441 ($\sim0.2$~dex more enriched than the remainder of the $s$-rich group) discovered in \citetalias{montypaperI}. A review of nucleosythentic sites of the $s$-process is given in Sec.~3.1.4 in \citetalias{montypaperI} \citep[the classification of nucleosythentic groups follows the system set out by the seminal paper of][]{burbidge}. Briefly, the main site of the $s$-process, contributing predominately to the production of elements from Sr to Nd, is asymptotic giant branch stars \citep[AGBs,][]{busso99, karakas10, karakas14, chiaki20}.

In \citetalias{montypaperI}, the average chemical abundances measured in NGC~288 showed good agreement with the average abundances measured in the $s$-weak group. We also used two lines of evidence to infer that the $s$-rich group may be younger than the $s$-weak group (see Sec.~4 of \citetalias{montypaperI}). If we accept the $s$-weak group as the older and \textit{primordial} group in NGC~362, comparisons between the $s$-weak group and NGC~288 hinted at primordial similarities between the two clusters. We explore this idea further in the top panel of Fig.~\ref{fig:clusterdif1}, where an element-by-element comparison between both $s$-process group and NGC~288 is presented. In the panel, the open circles show the abundance difference between NGC~288 and the $s$-rich group in NGC~362, while the filled circles represent abundance differences between NGC~288 and the $s$-weak group in NGC~362.

To evaluate the differences we take a simple approach by calculating the average (differential) abundance for each well-measured element ($n$ lines > 2, with the exception of Eu and neglecting the light elements) in each cluster and then calculate the absolute difference. The error on the cluster-to-cluster difference is defined as the sum in quadrature of the average error in that element, in each cluster. The first thing to notice is the confirmation of the conclusion from \citetalias{montypaperI} that the $s$-weak group and NGC~288 show greater similarity than the $s$-rich group and NGC~288 (13 of the 20 elements show greater similarity as shown in Fig.~\ref{fig:clusterdif1}).

The heavy $s$-process abundances (Ba and heavier) within the two ``primordial'' populations is strikingly similar, as well as the light elements (particularly Si through to Ti). The most prominent differences between the two primordial populations are seen in the Fe-peak elements, Cr, Fe and Co, the light $s$-process elements Y and Zr and the $r$-process elements Nd and Eu \citep[created via neutron star mergers and magneto-rotational supernovae][]{cote18, chiaki20}. Comparing the $s$-rich group in NGC~362 with NGC~288, reveals the same differences between elements as the $s$-weak/NGC~288 comparison, except the scale of the differences is larger overall.

To explore how chemically distinct the two GCs are relative to a GC that is \textit{not} dynamically associated with the GSE, we compare our cluster abundances to the disc-like GC, NGC~6752. The bottom panel in Fig.~\ref{fig:clusterdif1} shows the absolute differences between our two GCs and NGC~6752 using the average cluster abundances (considering all the stars in NGC~362, except for the $s$-process enhanced star) and data for NGC~6752 from the differential study of \citet{yong2013}. NGC~6752 is selected for comparison as it is one of the few GCs examined using high precision differential abundance analysis \citep[the chemically complex GC, M~22 being the only other,][]{mckenzie22} and has a similar metallicity to NGC~288/NGC~362 \citep[$\mathrm{[Fe/H]}\sim-1.48$, as calculated using NLTE,][]{kovalev}. Additionally, we use the same reference star in \citetalias{montypaperI} as \citet{yong2013} (mg9), therefore the two studies are anchored to the same zero-point. 

\todo{We are cautious about discussing the origin of NGC~6752 as it has disc-like kinematics but is chemically complex \citep{yong2013}. Furthermore, it has been dynamically associated with both the MW disc \citep[implying an in-situ origin, ][]{massari19} and with the \textit{Kraken} merger event \citep{callingham22}. Regardless of the origin of NGC~6752, it has not yet been associated with GSE and thus we treat it as a cluster with a distinctly different origin to both NGC~288 and NGC~362. Examining the bottom panel of Fig.~\ref{fig:clusterdif1}, both clusters show significantly different chemistry relative to NGC~6752 for every element in common. The largest differences between both GCs and NGC~6752 are seen in the heavier elements.}

Finally, as a measure of how similar our two GCs are relative to the chemical similarities seen in \textit{known} galactic siblings, we turn to the Large Magellanic Cloud (LMC) GC system. We employ data from the two studies of \citet{mucciarellithreegcs} and \citet{mucciarellibloodties} to explore the differences between a set of three similar metallicity LMC GCs (NGC~1786: [Fe/H]$=-1.75$, NGC~2210: [Fe/H]$=-1.65$ and NGC~2257: [Fe/H]$=-1.95$) and the differences seen between the binary LMC GC system composed of NGC~2136 ([Fe/H]$=-0.40$) and NGC~2137 ([Fe/H]$=-0.39$), respectively. In the case of the binary GC system, \citet{mucciarellibloodties} suggest from the level of chemical similarity that the GCs likely formed from the fragmentation of the same molecular cloud. Both studies perform the analysis in a homogeneous way, avoiding systematic offsets introduced when combining data from independent studies. 

\begin{figure}
    \centering
    \includegraphics[width=\linewidth]{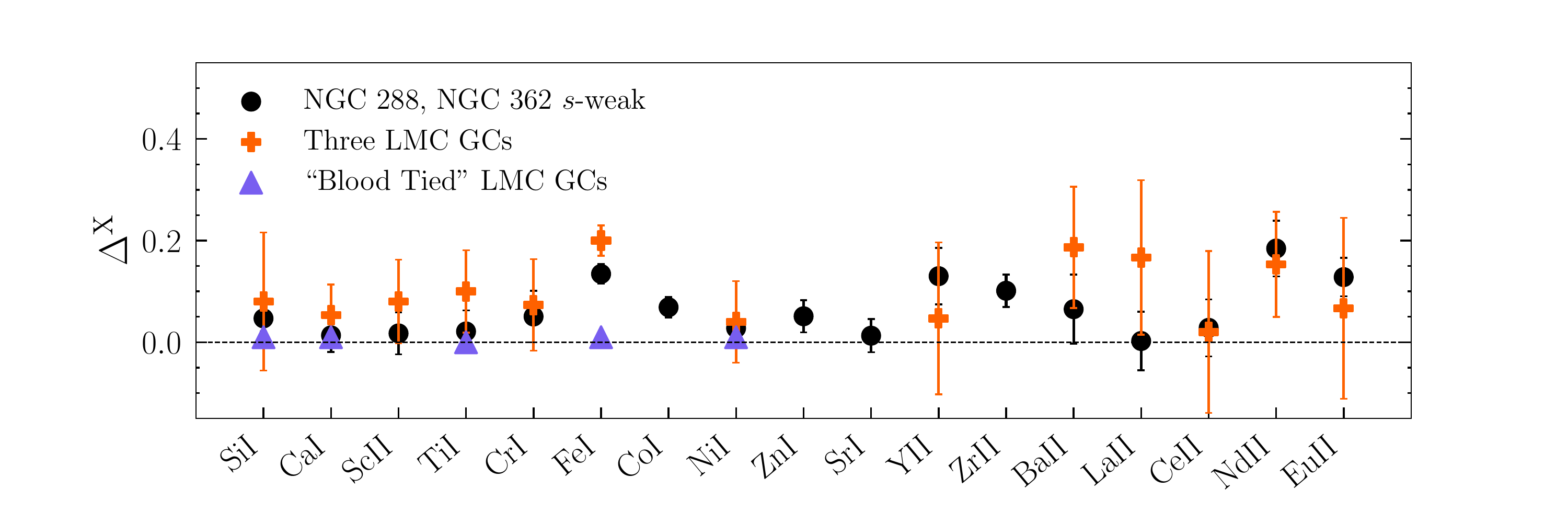}
    \includegraphics[width=\linewidth]{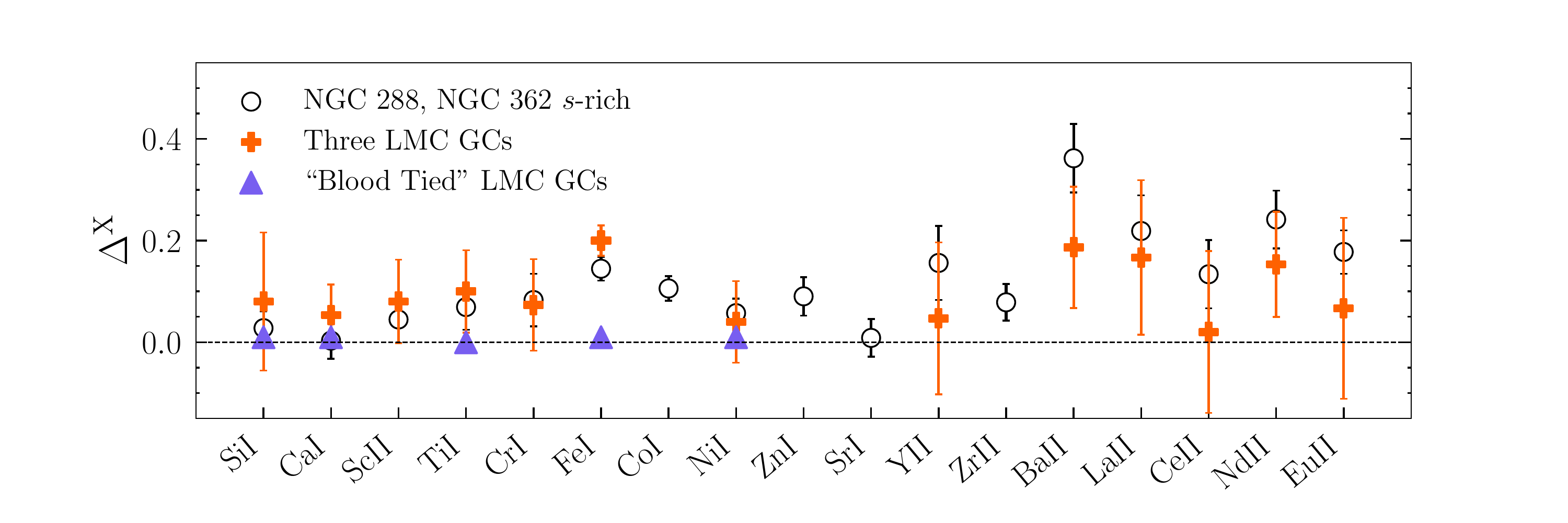}
    \caption{Comparison of the average abundance differences between NGC~288 and the two $s$-process groups in NGC~362 (\textit{top:} $s$-weak, \textit{bottom:} $s$-rich) to the average differences between the three LMC GCs, NGC~1786, NGC~2210 and NGC~2257 (orange) and the binary (``blood tied'') LMC GCs, NGC~2136 and NGC~2137 (purple) from the studies of \citet{mucciarellithreegcs} and \citet{mucciarellibloodties} respectively.}
    \label{fig:clusterdif2}
\end{figure}

Fig.~\ref{fig:clusterdif2} shows the average differences between each set of GCs, in each of the three studies. \todo{The intention of this comparison is to establish a benchmark for how similar GCs associated with a single parent galaxy can be\footnote{Note that while we assume the three LMC GCs we examine are true galactic siblings which formed alongside the LMC, the LMC is proposed to host at least one accreted GC \citep[NGC~2005,][]{mucciarelli21}. We avoid examining NGC~2005 for this reason.}. Therefore, at no point do we perform a direct comparison between our MW GCs and the LMC GCs.} The average abundance differences in the set of three LMC GCs are shown in orange in both panels of Fig.~\ref{fig:clusterdif2}, while the differences in the two GCs with ``blood ties'' are shown as the purple triangles. Note that the average abundances for the three LMCs are taken from the set of all three possible combinations. The error bars are calculated in the same way as Fig.~\ref{fig:clusterdif1}

As mentioned in the previous paragraph, given that the three LMC GCs, NGC~1786, NGC~2210 and NGC~2257 are suspected siblings, their chemical similarity will inform our expectations of the sibling scenario for NGC~288 and NGC~362. The definite siblings (i.e. NGC~2136 and NGC~2137) serve as an extreme example for how similar sibling GCs can be and provide an upper limit on the level of chemical similarity between GCs. For the three LMC GC siblings we find weighted average differences in the $\alpha$-elements Si, Ca and Ti of $0.07\pm0.02$~dex, $0.12\pm0.06$~dex in the Fe-peak elements Sc, Cr, Fe and Ni, $0.12\pm0.07$~dex in the $s$-process elements Y, Ba, La and Ce and a difference of $0.12\pm0.04$~dex in the $r$-process elements Nd and Eu.

Comparing the three LMC GC siblings with NGC 288 and the $s$-weak group in NGC 362 (top panel of Fig.~\ref{fig:clusterdif2}), one can see that our two GCs generally show greater chemical similarity than the three LMC GCs. The only exception to this is in the heavy elements Y, Nd and Eu. In the case of Nd and Eu, the two elements have a considerable contribution from the $r$-process \citep[Nd:~$\sim43\%$, Eu:~$94\%$,][]{bisterzo11} suggesting this could be the source of the disagreement. 

This exception is repeated in the lower panel of Fig.~\ref{fig:clusterdif2}, where the differences between the three LMC GCs are again shown alongside the differences between NCG~288 and the $s$-rich group in NGC~362. Again, our two GCs show greater agreement across most elements except the heavy elements. In this case, our GCs show greater differences in all the elements heavier than Sr. With the exception of the disagreement in $r$-process enhancement between NGC~288 and NGC~362, the primordial populations in our GCs (NGC~362 $s$-weak) show greater agreement overall when compared to similarities seen in the three LMC sibling GCs. As expected, the LMC GC twins, NGC~2136 and NGC~2137 show the smallest disagreements. 

\todo{\emph{Our second conclusion is that NGC~288 and NGC~362 display comparable, if not greater chemical similarities to those seen in the three assumed galactic siblings, the LMC GCs NGC~1786, NGC~2210 and NGC~2257. The only exception to this is in the heavy $s$- and $r$-process elements. We interpret this as support for the galactic sibling scenario.}}

\subsection{Accretion Scenario II: Complete Strangers}
Approaching the accretion scenario for NGC~288 and NGC~362 from the other perspective, we now present primarily chemical evidence that the GCs are not galactic siblings and were not accreted together as part of the GSE event. As discussed in Sec.~\ref{sec:intro}, this is in agreement with the findings of \citet{myeong18, myeong19}, who tentatively tag NGC~362 to the Gaia-Sausage accretion event. This is also in-line with the findings of \cite{sun23}, who dissect GSE into four dynamical groups, proposing that GSE is composed of an additional three unique accretion events. They assign NGC~288 to ``GSE-b'' while leaving the association of NGC~362 to the main GSE progenitor unchanged. Because the Sausage event occupies a smaller region of $E$-$L_{z}$ space, due to it's nearly net zero angular momentum, fewer GCs are declared to have a strong association. In the following sections we discuss the intrinsic dynamical differences between the GCs, exploring potential explanations, and present the chemical distinctions between the GCs.

\begin{figure}
    \centering
    \includegraphics[scale=0.195]{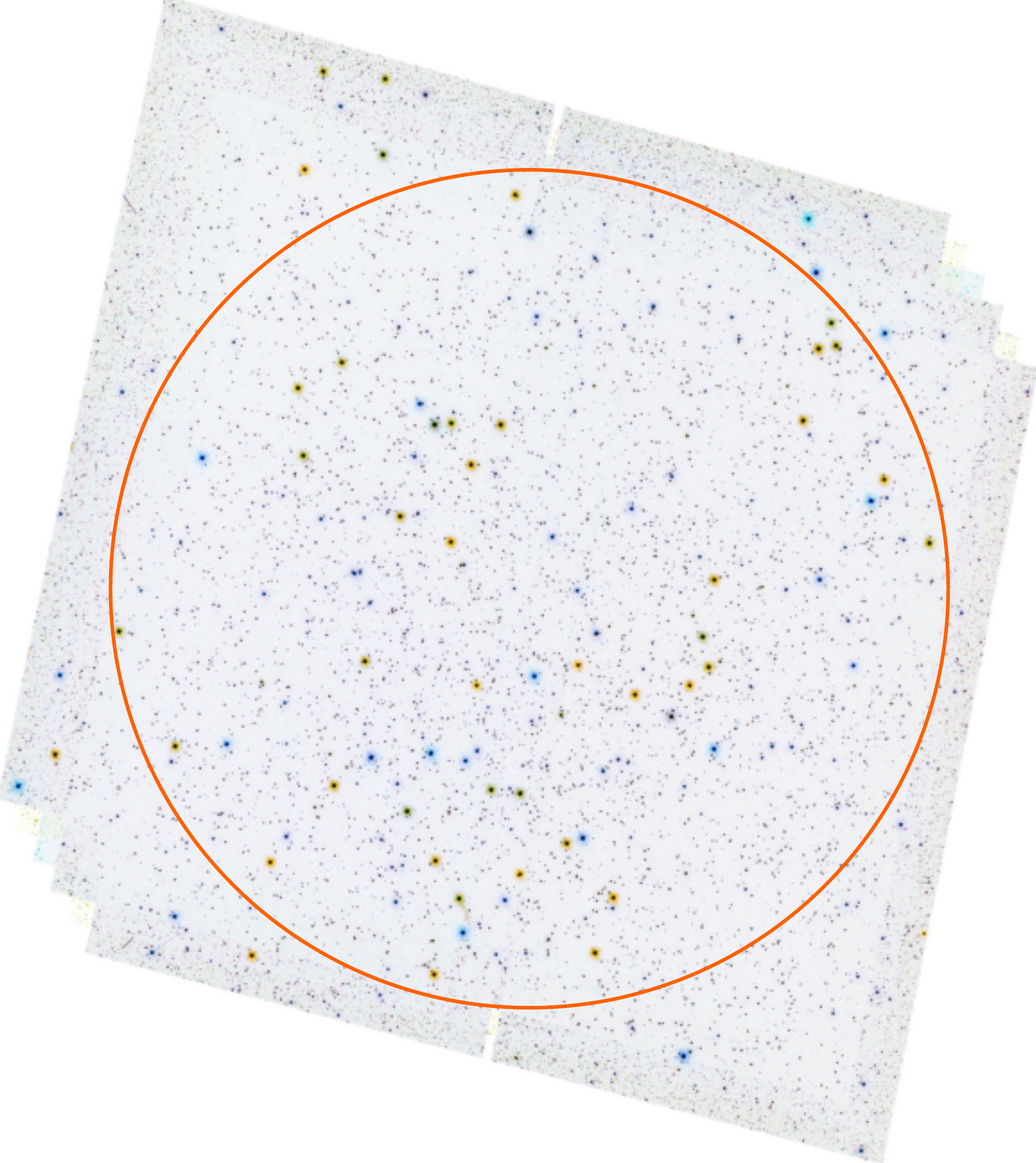}
    \includegraphics[scale=0.195]{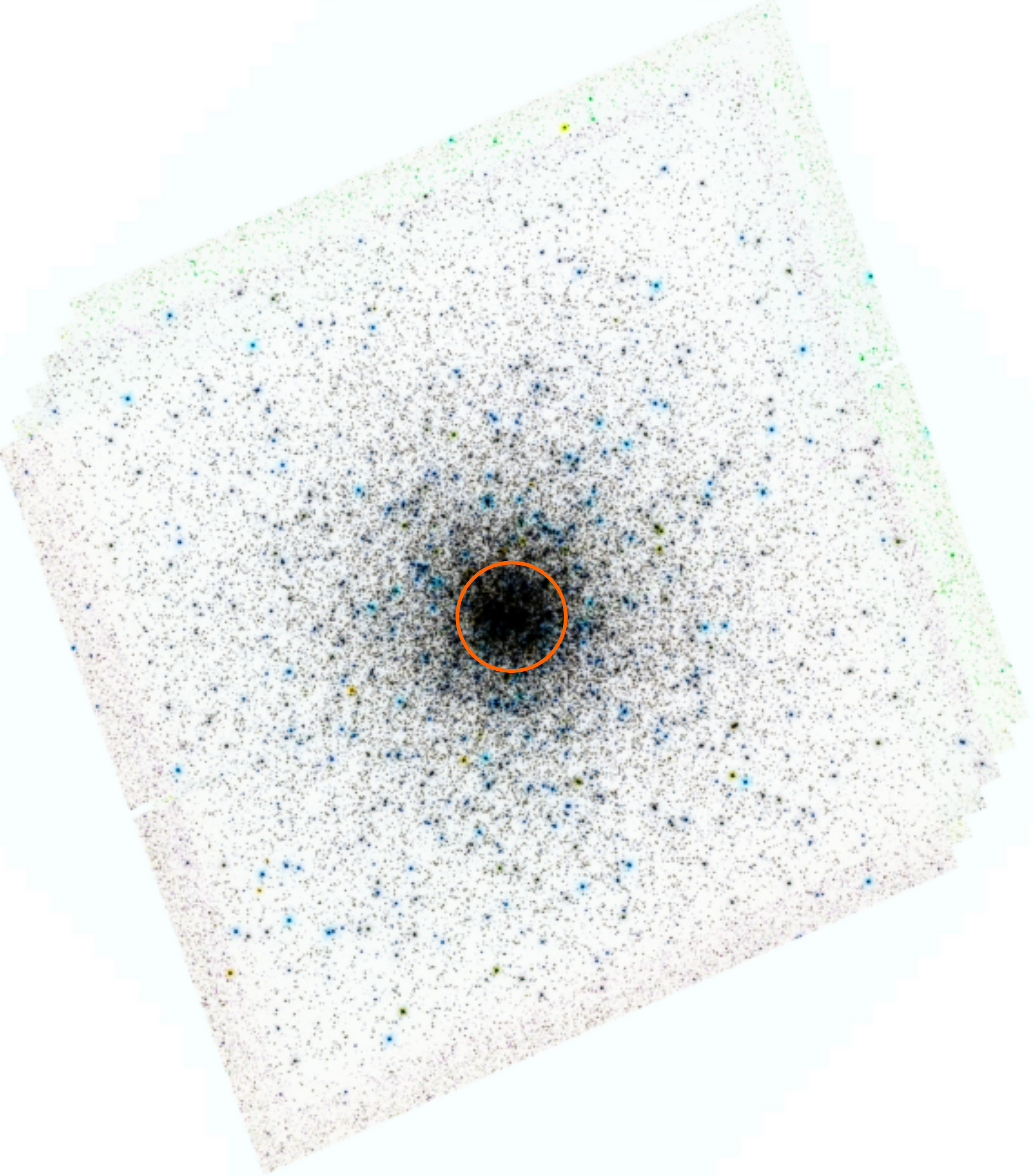}
    \caption{Snapshots from the HST HUGS Survey \citep{piottohugs, hugssurvey2} of NGC~288 (\textit{left}) and NGC~362 (\textit{right}) using the HST F275W band, the cluster core radii are marked in each image as the orange circle. The two images have similar total exposure times (NGC~288: 400s, NGC~362: 519s). Note the stark difference in the physical scale of the cluster core radii and recall that the clusters are at approximately the same \todo{heliocentric} distance \citep[$\sim9$~kpc][]{baumgardtdist}. \todo{Both GCs are found near the radially solar circle and below the MW disc, placing them on halo-like orbits.}}
    \label{fig:gcsnaps}
\end{figure}

\subsubsection{Dynamical Evidence Against a Common Origin}
\label{sec:dynagainst}
In the case of NGC~288 and NGC~362, one needs to do little more than look at images of the two clusters (Fig.~\ref{fig:gcsnaps}) to conclude that they have either, i) had significantly different dynamical histories, or ii) were born with very different initial conditions. Addressing the first question, as discussed in Sec.~\ref{sec:dynsib} the two GCs share many orbital characteristics in common. Despite this, they have vastly different internal characteristics. For example, the half-mass radius of NGC~288 is significantly more extended than NGC~362 \citep[by a factor of three,][]{baumgardthilker}. NGC~288 is also less massive than NGC~362 with a present day mass of $9.34\times10^{4}$~M$_{\odot}$ vs. $2.84\times10^{5}$~M$_{\odot}$ \citep{baumgardthilker}, resulting in a significantly lower density overall. This is readily apparent in Fig.~\ref{fig:gcsnaps}, where HST images of the two clusters are shown (NGC~288 on the left and NGC~362 on the right). The core radii of the two GCs are marked with orange circles. Note that the clusters have approximately identical heliocentric distances \citep{baumgardt21}. 

\todo{In the following paragraphs we explore two main questions related to the dynamical histories of the clusters and the hypothesis posed at the beginning of this section. The first, is that given the upper bound on the time of accretion of the Sausage merger \citep[10~Gyr,][]{gallart19, bonaca20, fattahi19, borre22}, could both clusters have survived to present day on their current orbits, or would they have been tidally disrupted by the MW. The second question is whether different conditions at birth could explain the present-day cluster characteristics. This question is motivated by the vastly different densities and yet nearly identical tidal radii the two clusters \citep[as calculated according to eq. 8 of][]{webb13}.}

One way to investigate the question of cluster disruption is to look at the timescales of dynamical processes within the two clusters. Within the cluster, internal two-body relaxation coupled with external disc and bulge shocking drives a redistribution of energy. This leads to mass segregation on timescales dependent on the magnitude of the interactions. In turn, tidal interactions with the MW gradually deplete the outer parts of the cluster resulting in a halo of potential escaping stars \citep{ostrikerchev, gnedin97, heggie03}. Considering only two-body relaxation and under particular assumptions, the relaxation time at the half-mass radius of a GC can be estimated \citep{spitzer69, galdyn}. In the case of NGC~288 ($r_{\mathrm{hm}}=8.37$~pc), the half-mass relaxation time is 2.95 Gyr, while in NGC~362 ($r_{\mathrm{hm}}=3.79$~pc) the relaxation time is somewhat shorter at 1.45 Gyr, reflecting the differences in cluster densities \citep{baumgardthilker}. 

As mentioned, in addition to two-body relaxation, disc and bulge shocking must be considered for the two GCs as they have experienced a significant number of disc passages over the last 10~Gyr. Given their present-day orbits many of these passages occurred in regions of high disc density. Disc shocking timescales can be estimated through arguments of energy injection timescales and binding energy at different points in the cluster \citep[e.g. Eqn. 12.9 in][]{heggie03} but is most accurately calculated via \textit{N}-Body simulations. \textit{N}-Body simulations of the two clusters were performed by \citet{baumgardt19} who predicted the dissolution times \todo{integrating forward from \textit{present day}} conditions. They found a dissolution time of 10.2~Gyr for NGC~288 and 10.8~Gyr for NGC~362. Given that they are expected to survive for another $\sim10$~Gyr, one can conclude that they have survived for at least that long already - answering the first question posed at the beginning of this section.

\begin{figure}
	\includegraphics[width=\linewidth]{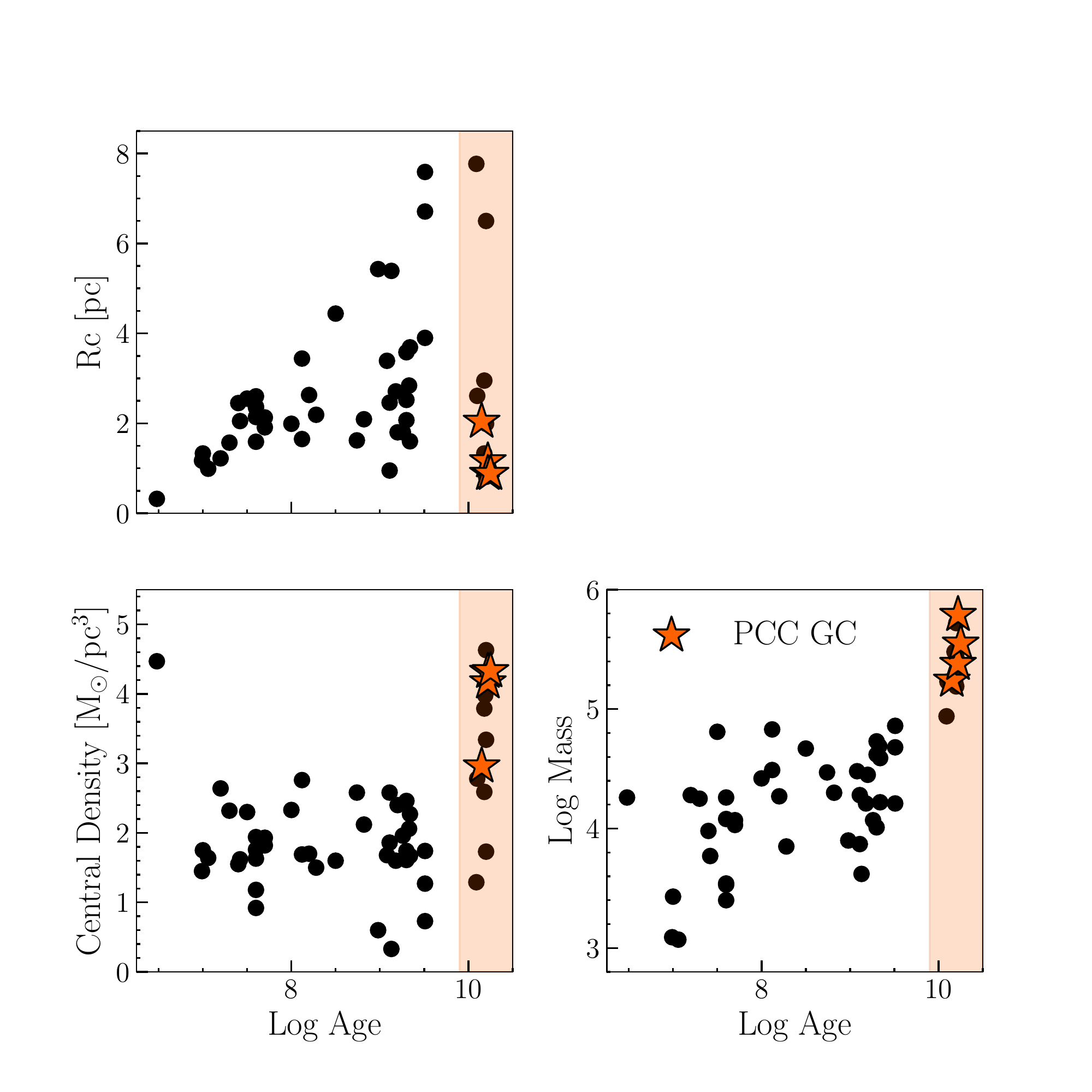}
	\caption{Core radii (\textit{top}), log mass (\textit{bottom left}) and \todo{central density (\textit{bottom right})} as functions of age for LMC GCs from \citet{mackey03}. Proposed PCC clusters are shown as the orange stars. The orange shaded region marks the area where NGC~288 and NCG~362 would occupy given their ages, and initial and present day masses. Note that both PCC and non-PCC clusters are found in this region.}
    \label{fig:lmc_rcore}
\end{figure}

The second question posed in this section addresses the initial conditions of the clusters, could this explain how drastically different they are today? To help with our discussion, we start with an estimate of the initial cluster masses. According to \citet{baumgardt19} utilising the same simulations that determined the lifetime of the GCs, the initial cluster masses were estimated to be $4.1\times10^{5}$~M$_{\odot}$ \todo{(NGC~288)}
and $1.4\times10^{6}$~M$_{\odot}$ \todo{(NGC~362)}. Given these initial masses and the present day cluster masses \citep{baumgardthilker}, both clusters have lost approximately 75\% of their initial mass. \todo{Note that this estimation ignores the effects of tidal shocks via interactions with giant molecular clouds during cluster formation. Therefore they represent the initial masses post-shock phase and provide a lower limit on mass-loss.} 

Another interesting thing to note is that NGC~362 is proposed to be a ``post-core-collapsed'' (PCC) cluster \citep{dalessandro13, libralato18}. Core collapse indicates that runaway mass segregation (driven primary by two-body relaxation) has occurred within the cluster resulting in the central density reaching a maximum. To assess the diversity of characteristics for clusters still found in-situ around their parent galaxy, we turn to the GC system of the LMC utilising the results of \citet{mackey03}. 

\todo{Three relevant trends to our discussion are shown in Fig.~\ref{fig:lmc_rcore}, where the core radius, mass and central density of LMC clusters from \citet{mackey03} are plotted against age.} Clusters that they believe are definitely PCC (and those speculated to be) are shown using the orange starred symbols in both panels. Analogues of NGC~288 and NGC~362 (sharing similar present-day characteristics) can be seen in the two panels among the oldest LMC GCs highlighted within the orange shaded regions. \todo{Three features are apparent, i) clusters with similar ages can show a variety of core radii, ii) only the most massive of the ancient LMC clusters have survived to present day and iii) the central density of the surviving GCs spans a large range of values} 

This suggests that clusters born in the same system can survive to the present day \textit{and} display significantly different internal characteristics \citep[assuming they all formed in-situ, unlike the proposed scenario for NGC~2005,][]{mucciarelli21}. This conclusion is supported both by simulations of GC formation like those of \citet{mckenzie21} who saw physically diverse GCs form with similar metallicites and by $N$-body simulations that reconstruct the initial cluster masses \citep[e.g.][]{baumgardthilker}. 

To further validate this idea, we explored the distribution of GC core radii as a function of radial distances from the optical centre of the LMC. Among the oldest clusters, both PCC and non-PCC GCs were found at similar radial distances. Although this is only a snapshot of the cluster orbits and may not be indicative of the average radial distances, it does reinforce the idea that clusters at similar locations within a galaxy can show very different internal characteristics. Given both the initial and present day masses of both NGC~362 and NGC~288 and recent estimates of their ages \citep[$\sim11$~Gyr,][]{vandenberg13}, both clusters would be coincident with the clump of GCs shown in the upper right corner of the right panel in Fig.~\ref{fig:lmc_rcore}, suggesting that at least in the environment of the LMC they would have survived until present day.

\todo{\emph{Our third conclusion is that there is no strong dynamical evidence against a common origin for the two clusters. The drastically different present-day characteristics of NGC~288 and NGC~362 can be explained if they formed with significantly different initial conditions. This is supported by the diversity of ancient clusters within the LMC.}}

\subsubsection{Chemical Evidence Against a Common Origin}
\label{sec:chemevidagainst}
To explore the chemical differences between the two clusters that could exclude the sibling scenario, we compare the average cluster abundances with trends found in GSE and Sausage stars. In this comparison, if either NGC~288 or NGC~362 is found to be chemically distinct from the GSE, then this provides evidence against the common origin hypothesis. A comparison is possible thanks to the numerous studies examining the chemistry of GSE and Sausage stars (see Sec.~\ref{sec:intro}). Two recent studies by \citet{matsuno21} and \citet{aguado21} have published dedicated high precision chemistry of stars tagged to the two proposed accretion scenarios.

Fig.~\ref{fig:clusterdif1} reveals that both $s$-process populations in NGC~362 show similar disagreement in the abundance of Eu when compared to the average value in NGC~288. The other element that shows the largest disagreement between the $s$-rich population and NGC~288 is Ba. Given the dominance of the $s$- and $r$-process production pathways for Ba and Eu respectively \citep{bisterzo11}, we select these two elements to act as characteristic elements representative of their nucleosynthetic groups and choose to compare the two with measured GSE and Sausage abundances.

Ba abundances in local group dGals vary, with some dGals showing Ba-enhancement relative to MW field stars at the same metallicity \citep[e.g. Fornax][]{shetrone03, tolstoy, letarte10, lemasle14}, while others show MW-like Ba abundances \citep[e.g. Sculptor][]{shetrone03, tolstoy, hill19, skull19}. Unfortunately, both Sausage and GSE stars appear to occupy the latter distribution \citep{monty20, aguado21, matsuno21}, making [Ba/Fe] alone an uninformative discriminator of galaxy membership in this case. 

The ratio of Ba to Eu serves as a diagnostic of the relative contributions of $s-$ and $r$-processes within systems, with dGals ratios favouring $r$-process dominance at early times \citep[][and references therein]{tolstoy}. Both Ba and Eu abundances were measured in GSE \citep{matsuno21} and Sausage stars \citep{aguado21} using data from the GALAH Survey \citep[$R\sim28,000$,][]{galahdr3} and observations from VLT/UVES respectively. Both studies found that the [Ba/Eu] abundance ratios in their stars pointed to a strong dominance in $r$-process sources due to an enhancement in Eu. This was also hinted at earlier in the study of \citet{yuan20}, where they recovered two known Eu-enhanced stars ([Eu/Fe]$>+1$) in GSE through dynamical tagging. 

We find this is also the case in NGC~288 and NGC~362 where we detect a marginal $r$-process enhancement relative to $s$-process, with negative [Ba/Eu] values in both clusters (in the case of the primordial $s$-weak group in NGC~362). Average differential ratios of Ba to Eu compared to the average measurement error (adding the measurement errors of Ba and Eu in quadrature) were found to be $\Delta^{\mathrm{Ba/Eu}}=-0.10\pm0.04$ in NGC~288 and $\Delta^{\mathrm{Ba/Eu}}=-0.16\pm0.07$ in the $s$-weak group. \todo{Recall that $\Delta^{\mathrm{X/Y}}$ in this case refers to the differential abundances for the two clusters derived in \citetalias{montypaperI}.} When considering all the $s$-process elements with multiple line measurements (Y, Ba, Ce and La) vs. the $r$-process element Eu, the average ratio of $s$/$r$ in NGC~288 is $-0.05\pm0.04$~dex and $-0.19\pm0.05$~dex in the $s$-weak group in NGC~362. 

The greatest $s$/$r$ deficit in each cluster or group, was found in Ba in NGC~288 ($\Delta^{\mathrm{Ba/Eu}}=-0.10\pm0.04$) and in Y in both $s$-process groups NGC~362 ($\Delta^{\mathrm{Y/Eu}}=-0.26\pm0.05$  in the $s$-weak group.) Additionally, the average \todo{differential} Eu abundances in the both clusters is elevated, with NGC~362 showing greater Eu-enhancement than NGC~288 by $\sim0.15$~dex. Note that the Eu-enhancement does not change between groups in NGC~362, leading to the conclusion in \citetalias{montypaperI} that the enhancement in the $r$-process element is primordial.

Another useful ratio to explore is [Eu/Mg] which is sensitive to the delay time of neutron star mergers in the production of Eu. \citet{matsuno21} found that [Eu/Mg] appeared significantly higher in GSE stars when compared to a population of in-situ MW stars ([Eu/Mg]~$\sim0.40$~dex in GSE stars vs. [Eu/Mg]~$\sim0.20$~dex in in-situ stars at [Fe/H]~$\sim-1.3$~dex). Although we do not consider Mg to be one of the best-measured elements in our study, we consider the cluster abundances of [Eu/Mg]. Both groups in NGC~362 show higher $\Delta^{\mathrm{Eu/Mg}}$ abundance ratios than NGC~288, with an average difference of $\Delta^{\mathrm{Eu/Mg}}=0.27\pm0.07$ in the $s$-weak group compared to $\Delta^{\mathrm{Eu/Mg}}=0.19\pm0.11$ in NGC~288. The ratio of [Eu/Mg] in both groups in NGC~362 place the cluster between average GSE and in-situ MW values, while NGC~288 falls directly on-top of the average in-situ ratio.

Choosing a more reliable $\alpha$-element tracer in our study, namely Si, we see the same but more marked difference between the two clusters ($\Delta^{\mathrm{Eu/Si}}=0.27\pm0.04$ in the NGC~362 $s$-weak group and $\Delta^{\mathrm{Eu/Si}}=0.09\pm0.03$ in NGC~288). NGC~362 shows on average a $\sim0.2$~dex higher [Eu/Si] abundance ratio compared to NGC~288.

To confirm the difference in [Eu/$\alpha$] ratios between the two clusters, we cross-matched the catalogues of \citet{vasilievbaumgardt} and GALAH DR3 \citep{galahdr3} to select high probability cluster members from our two GCs within GALAH. The catalogues were then further refined by setting the DR3 flag \texttt{S/N\_c3\_iraf} > 30. \todo{In total we considered 23 stars in NGC~362 and 76 in NGC~288.} The average values of [Eu/Si] in the cleaned NGC~362 and NGC~288 catalogues were 0.50 and 0.26~dex respectively. NGC~288 showed a larger spread in [Eu/Si] than NGC~362, with $\sigma$ of 0.2~dex vs 0.1~dex in NGC~362 for a similar number of stars and range in metallicity. The difference in [Eu/Mg] between the two clusters is less clear in GALAH, but NGC~362 still displays a higher abundance ratio overall.
 
\begin{figure}
    \centering
	\includegraphics[scale=0.425]{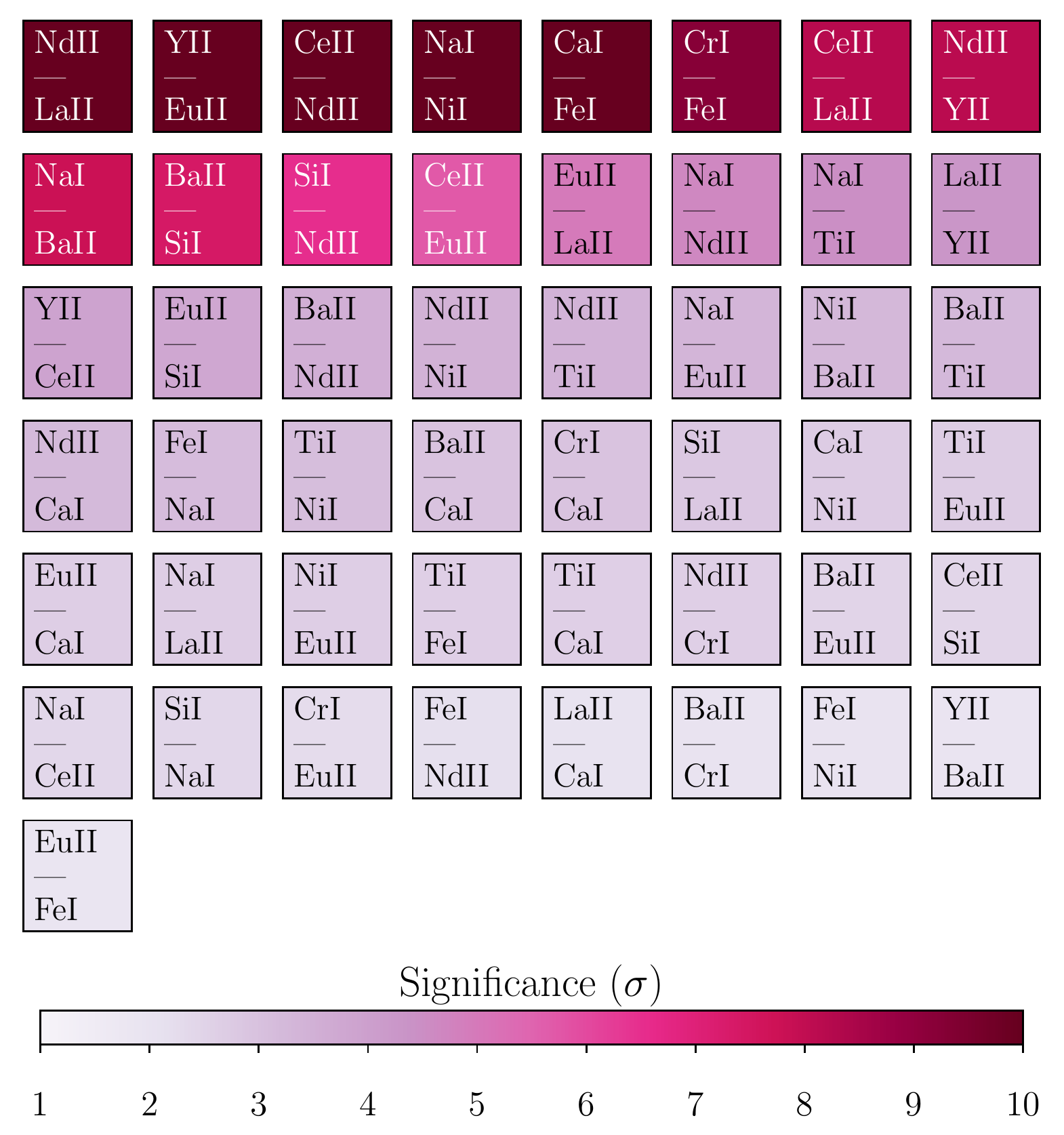}
	\caption{All possible well-measured element ratios coloured by how significantly different they are between the primordial $s$-weak group in NGC~362 and NGC~288. Only elements with more than one line are considered (with the exception of Eu are shown). Note the differences in the ratio of the first peak $s$-process element Y and the $r$-process element Eu between the clusters.}
    \label{fig:allsigdif}
\end{figure}

Neglecting a potential association with GSE, we also delve deeper into the abundance differences between the two clusters, comparing the primordial $s$-weak group in NGC~362 to all stars in NGC~288. The results of this are presented in Fig.~\ref{fig:allsigdif} where all possible well-measured element ratios are compared between \todo{the $s$-weak group in NGC~362 and all stars in NGC~288}. Each ratio is coloured by the significance, quantified as the ratio of the absolute difference between clusters over the uncertainty in the difference, only the elements with more than one line are considered (with the exception of Eu). Fig.~\ref{fig:allelemratios} shows the significance of the differences between all possible element ratios.

The first interesting difference in ratios we note is the ratio of Y to Eu, followed further down by the ratio of Y to Nd. These two ratios represent two measurements of the ratio of an $s$-process to $r$-process element in the clusters (Nd being both $s$ and $r$). This is in-line with the findings of Sec.~\ref{sec:chemsupport}, where the clusters show sibling-level similarities, with the exception of the heavy elements. Interestingly, the ratios of Ba to Eu are not significantly different, but the ratios of Zr to Nd (shown in Fig.~\ref{fig:allelemratios}) are. This could imply that the first $s$-peak elements, produced both by AGB and through core collapse supernovae (CCSNe), show greater distinctness between clusters vs. the AGB-dominated $s$-process elements like Ba \citep{hansen14, chiaki20}. In all cases, the ratios are significantly more negative ($\sim0.2$~dex) in NGC~362, further supporting the dominance of $r$-process production in the cluster.

\todo{\emph{Our fourth conclusion is that the abundance of Eu and ratio of [Eu/Si(Mg)] in the primordial $s$-weak group in NGC~362 more closely resembles Sausage and GSE stars at the same metallicity than NGC~288. Furthermore, distinct chemical differences are found between the two clusters, primarily in the ratios of the first peak $s$-process to $r$-process elements, with the $r$-process in NGC~362 showing clearer dominance.}}

\section{Fitting Nucleosynthetic Group Ratios using a Galactic Chemical Evolution Model}
\label{sec:gce}
To explore the chemical abundance histories of the host galaxy(ies) around which NGC~288 and NGC~362 formed, we now experiment with replicating the trends in both clusters using a galactic chemical evolution (GCE) model. This assumes the GC chemical abundances will reflect those of their host galaxy \todo{at the time of cluster formation}. We accept this assumption as we are not investigating GCE models with the intention of \textit{deriving} the dGal chemical evolution history. Instead, our aim is to explore whether the abundance ratios found in NGC~362 and NGC~288 can be explained using a \textit{single} model and whether this model is consistent with GSE, the proposed host galaxy. 

To support this assumption we reference several studies, the first being the study of \citet{ngc1261}, where the GSE-tagged GC NGC~1261 (tentatively tagged to the Sausage) was used to infer the $r$-process enrichment of GSE. This was done after confirming several abundance ratios in common between the GC and GSE stars. In their study of the LMC GC NGC~1718, \citet{sakari17} also recovered GC abundances consistent with LMC field stars. \cite{mcwilliam13} found the same in the Sagittarius dGal, where  Sagittarius field stars and tagged GCs showed consistent chemistry, particularly in $\alpha$-elements abundances. Further afield, the Fornax dGal GC H4 was also found to show chemical consistency with Fornax field stars in $\alpha$-abundances and clear chemical distinction from MW halo stars \citep{hendricks16}. Finally, in their study of chemical abundances in 15 LMC GCs derived from integrated light spectroscopy, \citet{chilinlmc} showed that the chemical evolution history of the LMC built using the GC abundances alone agreed with the history derived using resolved spectra of individual LMC stars. 

\subsection{The GCE Model}
\label{sec:describegcemodel}
We create our GCE model of GSE from the ``baseline'' model (Model A) of \citet[][discussed in Section 4 of their paper]{matsuno21}, which was created to both replicate and predict the chemical abundance ratios found in accreted GSE stars. To build our GCE we use the  \texttt{OMEGA} module \citep{omega} within the \texttt{NuPyCEE} framework \citep{nupycee}. \texttt{OMEGA} utilises both the NuGrid stellar evolution models \citep{nugrid1,nugrid2} and the \texttt{SYGMA} module \citep{sygma} to derive chemical yields associated with evolving simple stellar populations. We choose a closed-box model, primarily for simplicity and to avoid exploring galactic physics (i.e. in-flows and out-flows), which \citet{omega} found are not well-constrained when fitting a small number of abundance ratios. 

As a first step, we adopt approximately the same initial gas mass ($2.5\times10^{9}$~M$_{\odot}$), the same final stellar mass ($1\times10^{9}$~M$_{\odot}$) and the same total integration time (3~Gyr) as \citet{matsuno21} to build our model. Like \citet{matsuno21}, we also adopt a Chabrier IMF \citep{chabrier} with a lower mass bound of 0.1~M$_{\odot}$ and an upper bound of 100~M$_\odot$, adopt a power-law ($\beta-1$) for our SNIa delay time distribution\footnote{The delay time distribution captures the rate of events as a function of time following a burst of SF.} and assume an initial neutron star merger (NSM) rate efficiency of 0.5\% for stars in the mass range $8~\mathrm{M}{_\odot}<\mathrm{M}<20~\mathrm{M}_{\odot}$. 
Our baseline model differs from \citet{matsuno21} in that we apply mostly different yield tables and allow for the production of the Eu via multiple sources. For our AGB and massive star yields we make use of the yield table provided by \texttt{OMEGA}, which adopts yields from the MESA models only \citep{mesa} and fixes the electron fraction within neutron stars at $Y_{e}=0.4982$ following the prescription of \citet{young06}. For the SNIa yields, we adopt the same yields as \citet{matsuno21}, that of \citet{seitenzahl13}, using the table provided by \texttt{OMEGA} created to include a mixture of stable and unstable isotopic yields. Finally, we adopt the neutron star merger $r$-process yields of \citet{rosswog14}. 

As discussed in \citet{matsuno21}, their baseline GCE model is selected to explore the high [Eu/Mg] abundance ratios they observed in GSE stars in GALAH. To explain the elevated ratio (relative to MW field stars), they explore two possible explanations. The first is that Eu is overproduced in GSE due to a combination of a delayed production of $r$-process elements and prolonged SF history. The second is that Mg is under-produced as a result of a \textit{top-light} IMF (manifesting as less massive CCSNe and a decreased enrichment in Mg). Ultimately, their findings support the first scenario, that the overabundance of Eu is best explained by an excess of NSMs. This is broadly in agreement with previous work that suggests NSMs are the primary site of Eu production \citep{ji16, cote18}.

\citet{matsuno21} also discuss the implications of NSMs being the primary $r$-process site in GSE in the context of enrichment timescales and NSM delay times. Because GSE is a disrupted dGal, the time of accretion or disruption by the MW places a constraint on the end of SF within the dGal. As discussed in Sec.~\ref{sec:dynagainst}, estimates for the time of GSE-MW merger place an upper bound of $10$~Gyr ago (although it is hard to define when exactly accretion occurs and if this is coincident with the halting of SF). Unlike GSE, to explain the $r$-process enrichment patterns in the MW and surviving satellites, a secondary $r$-process enrichment site (in addition to NSMs) has been suggested \citep{coterproc, skul19, skularproc, chiaki20}. 

The necessity of a secondary formation site within the MW stems from the need to explain $r$-process enrichment at all times, coupled with the predicted delay time for NSMs. Estimates for the delay time range from 50~Myr to 4~Gyr when galactic chemical enrichment histories are used to place the constraints \citep{cote18, skularproc, naidu22, reyes22}, to $\sim7-14$~Gyr from host galaxy follow-up of the confirmed NSM, GW170817 \citep{blanchard17}. Assuming SF stopped in GSE $\sim10$~Gyr ago, the delay time for NSMs within the dGal is necessarily closer in agreement to the shorter delay times. While \citet{matsuno21} adopt a delay time of 20~Myr, \citet{naidu22} derive a minimum delay time of 500~Myr from additional spectroscopy of GSE stars, suggesting that rare CCSNe could explain some of the earliest enrichment in GSE and offering another $r$-process production site within the dGal.

\subsection{Replicating the chemical trends seen in both GSE stars and our clusters}
\label{sec:gceresults}
To test our baseline GCE model, we selected a representative element from four nucleosynthetic groups (discussed in-depth in \citetalias{montypaperI}). To select each representative element, we considered both the quality of the abundance measurements and the contribution of each nucleosynthetic process to the production of that element \citep{bisterzo11, chiaki20}. We chose Si to represent the $\alpha$-elements, Fe the Fe-peak elements and Ba and Eu to represent $s$- and $r$-process elements respectively. The quality of the fit between observations and model predictions for each of the six unique ratios was then assessed. 

\begin{figure}
	\includegraphics[width=\linewidth]{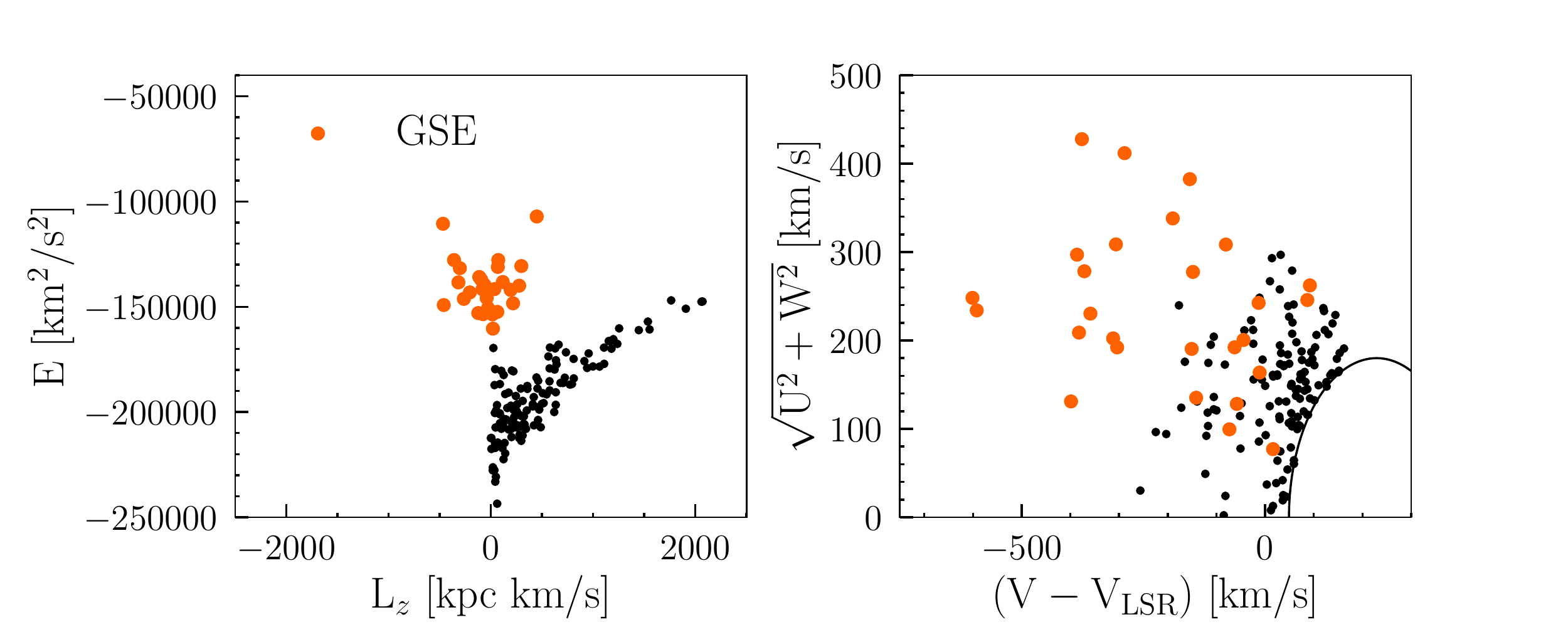}
	\caption{Selections of GSE stars from GALAH DR3 following several chemical, kinematic and dynamical cuts. The GSE selection is presented in orange, while a selection of in-situ stars is shown in black. The samples are shown in energy vs. z-angular momentum (\textit{left}) and in the classical Toomre diagram (\textit{right}). All orbital parameters are determined assuming a \texttt{McMillan17} potential in \texttt{galpy}.}
    \label{fig:insitugsegalah}
\end{figure}

To aid in assessing the quality of the GCE model, we i) selected the ratio of Eu to Si to act as the primary discriminator, based on the discussion presented in Sec.~\ref{sec:chemevidagainst} and ii) selected a sample of GSE stars from GALAH DR3 to better assess the quality of the fit. To select the GALAH GSE stars, we selected stars from Table 5 of \citet{buder22} and applied the following cuts,
\begin{enumerate}
    \item \texttt{probability\_chemical\_selection}~$\geq0.45$
    \item \texttt{probability\_dynamical\_selection}~$=1$
\end{enumerate}
%In-situ stars were selected via combining the GALAH DR3 \texttt{GALAH\_DR3\_main\_allstar\_v2} catalogue with the value added catalogues \texttt{GALAH\_DR3\_VAC\_GaiaEDR3\_v2}. 
Further cuts in the quality of the data (flags), kinematic and dynamical constraints were taken directly from Section 2 of \citet{matsuno21} and applied to the GSE population. All kinematic (UVW) and dynamical parameters (J$_{r}$, E and L$_{z}$) were determined assuming the \texttt{McMillan17} potential in \texttt{galpy} as in \citet{matsuno21}, integrating for 1~Gyr. The GSE sample was then further cleaned to only include giants with stellar parameters in the same range as our program stars (T$_{\mathrm{eff}}\in[3800, 4400]$~K, log~$g\in[0.2, 1.2]$~cm/s$^{2}$). The final sample selection of GSE stars is presented in Fig.~\ref{fig:insitugsegalah} showing the dynamical footprint of the sample in $E$ vs. $L_{z}$ space and in a classical Toomre diagram. In the diagram, we adopt v$_{\mathrm{LSR}}=229$~km/s \citep{eilers19}. 

Finally, as further verification of the GCE model, we used the Gaussian Mixture Model (GMM) of \citet{myeong22} to construct predictions for the mean GSE abundances in our six abundance ratios. In their study, \citet{myeong22} fit four distinct components to the MW halo within their model, GSE being the only accreted component. Because the GMM predictions for GSE occur at higher metallicities than our two GCs, we extended the predictions using linear fits. This was done using the angle between the semi-minor and semi-major axes of the covariance ellipse associated with GSE probability distribution functions. \todo{In the case of [$s$/$r$] the covariance ellipse is nearly circular and so the direction of the trend is highly uncertain. We consider the predictions from both the physically motivated (and tuneable) GCE model and the data-driven GMM to be complementary given the diversity of the two approaches. However, we acknowledge that the two approaches are not completely independent as GALAH data is used to some extent in both approaches to constrain the final model.}

To evaluate the baseline model and explore alternate solutions, we collapsed our two GCs (the two $s$-process groups in NGC~362, NGC~288) and the GALAH GSE data set into their average values with error bars denoting the intrinsic spread (standard deviation). The baseline model was then fit to the average ratio values over a metallicity range centred around GC averages ($-1.5<\mathrm{[Fe/H]}<-1.2$~dex), in the six different combinations. Initial evaluation of the baseline model found that model failed to reproduce the ratio of [Eu/Si] associated with both the GSE data points and GMM trends, underpredicting both. The only ratios in which the baseline model fit the GSE data points well was in [Fe/Eu], [Ba/Eu] and [Fe/Ba], suggesting the $s$- and Fe-peak elements were well-reproduced with the baseline model - but not the $r$-process abundances.

Given the failure of the baseline model to fit the GSE [Eu/Si] ratio (our primary discriminator), we decided to explore increasing the production of Eu in our models through varying two different parameters. The first parameter was the fraction of NSMs in the mass range $8~\mathrm{M}{_\odot}<\mathrm{M}<20~\mathrm{M}_{\odot}$, and the second was the transition mass from AGB to massive stars (the minimum mass for a CCSNe). Both should primarily effect the $\alpha$, $s$- and $r$-process channels. The remaining galaxy characteristics including the closed box assumption, yield tables and integration time remain unchanged as described in Sec.~\ref{sec:describegcemodel}

\begin{figure*}
	\includegraphics[scale=0.5]{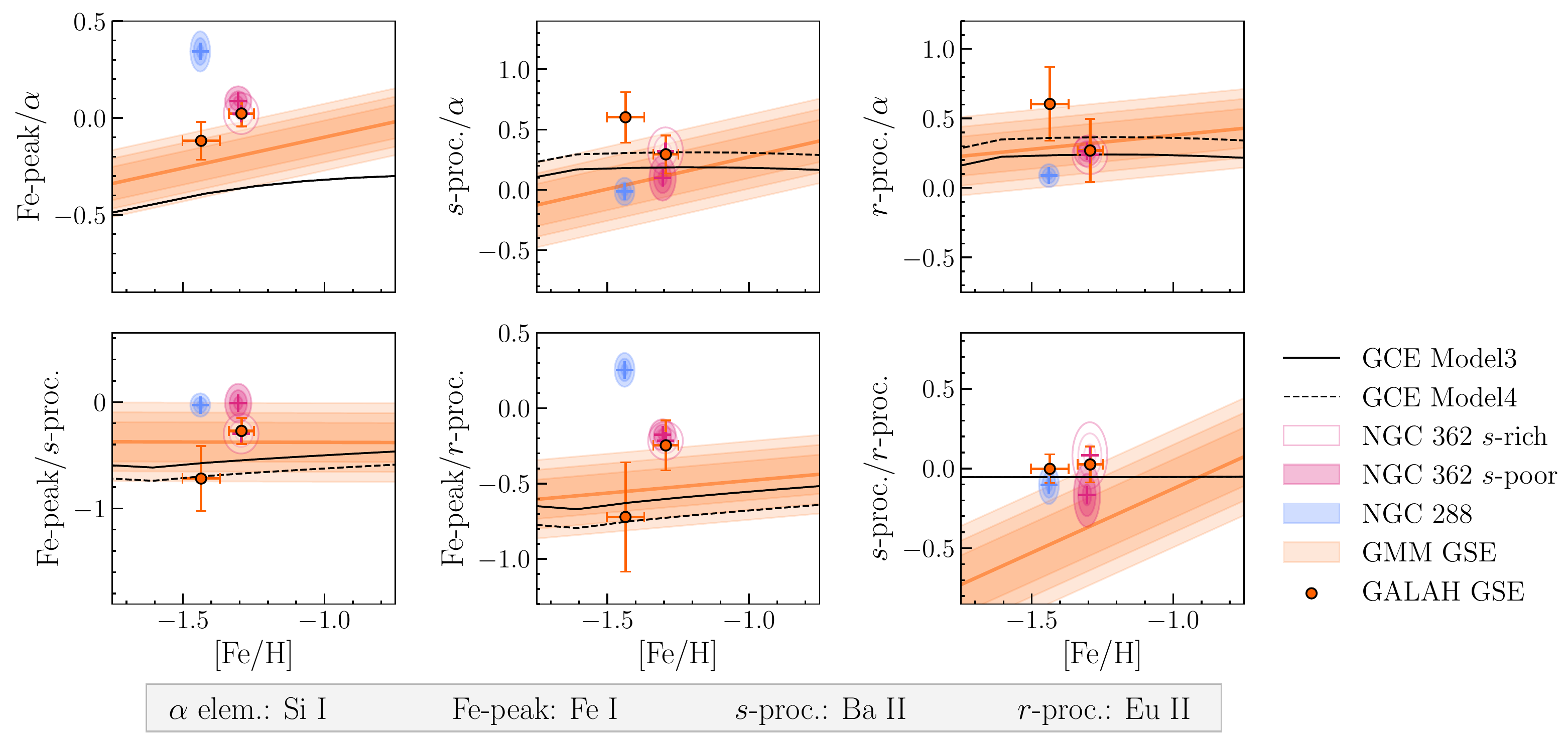}
	\caption{Predictions for the abundance ratios of four different nuclosynthetic channels (as characterised by the elements included in the key below the figure) from the two best-fit  galactic chemical evolution models discussed in Sec.~\ref{sec:gceresults}. Model parameters are given in Table~\ref{tab:gcemodels} and discussed in Sec.~\ref{sec:describegcemodel}. Average values for our two $s$-process groups in NGC~362, and a subset of GSE stars from GALAH are shown for each ratio. Trends for GSE predicted from the GMM of \citet{myeong22} are shown as the orange shaded regions marking the $1\sigma$, $1.5\sigma$ and $2\sigma$ deviations from the predicted trends. Note that the GCE models are overlapping in the panel showing the prediction for the ratio of Fe-peak/$\alpha$ and $s$/$r$.}
    \label{fig:gcemodel}
\end{figure*}

Because our aim was not to derive the GCE model for our two GCs, simply to explore how well a GCE model designed for GSE fits the data, we refrained from a full exploration of parameter space to derive a best-fit GCE model. Instead, we chose to simply modify the baseline model, exploring only transition masses (M$_{\mathrm{T}}$) in the range $8\mathrm{M}{_\odot}<\mathrm{M}<12\mathrm{M}_{\odot}$ and to gradually increase the NSM fraction (f$_{\mathrm{NSM}}$). This resulted in three more GCE models in addition to the baseline model, models 2, 3 and 4 with details provided in Table~\ref{tab:gcemodels}. From the predictions given for the three additional models, and considering our choice of emphasis on the fit to [Eu/Si], we found Model3 best-fit the GSE stars at higher metallicities (near that of NGC~362) and Model4 to best-fit the lower metallicities (near that of NGC~288). As listed in Table~\ref{tab:gcemodels}, both Model3 and 4 have a higher fraction of NSMs compared to the baseline model (0.25\% and 0.5\% respectively) and a higher turnover mass marking the change from AGB to massive star yields (10 and 12~M$_{\odot}$ respectively). The predictions from the two models are shown in Fig.~\ref{fig:gcemodel}. Model3 is shown using the solid black line and Model~4 the dashed line.

\subsection{Evaluating the GCE Model Fits to our GCs}
Regarding the fit to our GCs presented in Fig.~\ref{fig:gcemodel}, without considering the GCE models themselves, both groups in NGC~362 appear more similar to the average GALAH GSE abundances at the same metallicity. In fact, complete overlap is seen between the GALAH GSE average and one or both groups in NGC~362 in all six ratios. This was discussed to some extent already in Sec.~\ref{sec:chemevidagainst}. Also, one or both groups in NGC~362 appear within $2\sigma$ of the GMM predictions in every ratio, while this is only true for NGC~288 in four of the six ratios.  Naturally, this results in the GCE models better fitting the average abundance ratios in NGC~362 overall vs. NGC~288. This is especially apparent in the ratios of [$s$/$\alpha$] and [$r$/$\alpha$]. \todo{Note that the apparent disagreement between the GMM and GALAH data points in these ratios is likely due to different samples of GALAH GSE stars being selected by our study and that of \citet{myeong22}.}

Interestingly, Model3 with the lower fraction of NSMs and closer to the baseline model, better reproduces the abundances of both groups in NGC~362 than it does GSE. The abundance ratios in NGC~288 cannot be easily reproduced with a single model and instead vary between the models 3 and 4 depending on the ratio. In the case of the $s$-/$r$-process abundance ratio, both models predict the same ratio and are able to reproduce the ratio found in all groups equally. \todo{Finally, if we use the predicted elemental dispersion in the GSE GMM (the shaded regions in Fig.~\ref{fig:gcemodel}) as a proxy for the chemical (in-)homogeneity in the GSE progenitor, we can see that the disagreements between NGC~288 and GSE stars cannot be explained by chemical in-homogeneity alone. That is, if the GCs are truly siblings, this demands a degree of chemical in-homogeneity in the GSE progenitor that is larger than what is presently predicted by the GMM.}

\todo{\emph{Our fifth, and final conclusion, is that the abundance ratios in NGC~362 are well reproduced by a slightly modified GCE representative of GSE. It is unclear whether this is true in NGC~288. This supports a scenario in which the GCs are not siblings, or demands a greater degree of chemical in-homogeneity in the GSE progenitor than what is presently seen in the data.}}

\begin{table}
	\centering
	\caption{Galactic chemical evolution models of GSE created to explore fitting the abundance ratios found in our GCs. Each model is characterised by the two parameters we chose to tune; the transition mass from AGB to massive stars (M$_{\mathrm{T}}$) and the fraction of NSMs for stars in the mass range $8\mathrm~{M}{_\odot}<\mathrm{M}<12~\mathrm{M}_{\odot}$. Further details of the models are given in the text in Sec~\ref{sec:describegcemodel}. Predictions for the best-fit models, Model3 and 4 in the six abundance ratios are shown in Fig.s~\ref{fig:gcemodel}.}
	\label{tab:gcemodels}
	\begin{tabular}{lll} % four columns, alignment for each
		\hline
		Name & M$_{\mathrm{T}}$ [M$_{\odot}$] & f$_{\mathrm{NSM}}$\\
		\hline
		Model1 (Baseline)   & 8 & 0.005\\
		Model2  & 10    & 0.0075 \\
		Model3  & 12    & 0.0075 \\
		Model4  & 12    & 0.010 \\
		\hline
	\end{tabular}
\end{table}

\section{Summary}
\label{sec:sum}
Using data from our recent differential chemical abundance study of the MW globular clusters (GCs) NGC~288 and NGC~362 \citep[][Paper I]{montypaperI}, we explored the chemo-dynamical likelihood that the two GCs are galactic siblings. The two GCs exhibit similar metallicites ([Fe/H]$_{\mathrm{ave}}=-1.44$ in NGC~288 and [Fe/H]$_{\mathrm{ave}}=-1.30$ in NGC~362 from \citetalias{montypaperI}) and nearly identical orbits ($r_{\mathrm{peri}}=0.25$~pc, $r_{\mathrm{apo}}=10.90$~pc, $e=0.96$ in the case of NGC~288 and $r_{\mathrm{peri}}=0.38$~pc, $r_{\mathrm{apo}}=13.17$~pc, $e=0.94$ in NGC~362). Thus far only the position on the MW GC age-metallicity relation and/or the integrals of motion have been used to either infer or refute siblinghood for the two GCs \citep{myeong18, myeong19, massari19, callingham22, malhan22a}. Both GCs are proposed to have been brought in as part of Gaia-Enceladus merger \citep{helmi18, massari19}, while only NGC~362 has been tagged to the Gaia-Sausage \citep{belokurov18, myeong18, myeong19}. As part of our exploration, we assessed chemical and dynamical evidence both \textit{for} and \textit{against} the galactic sibling scenario.

To support the galactic sibling scenario, we integrated the two GC orbits in a static MW-like potential which - contrary to all previous tagging studies of these GCs - contained a representation of the MW bar. We explored two different models of the bar. The first model was represented using the Dehnen Potential, as implemented in the \textsc{python}-based galactic dynamics package \texttt{galpy} \citep{dehnenbar, galpy}. The second was built using Ferrers Potential (also as implemented in \texttt{galpy}) which represented both the bar and the MW bulge. Both weak and strong models of the bar (as quantified by the bar strength, $A_{f}$) were explored and each GC was integrated for 5~Gyr. In the case of integration in the presence of both a weak and strong bar, no major changes to the two cluster orbits were found. The $z$-component of the angular momentum in the two clusters was found to change by no more than $\sim40$~km/s~kpc in NGC~288 and no more than $\sim20$ ~km/s~kpc in NGC~362. In both cases, the clusters retained their associated with Gaia-Enceladus as defined by \citet{massari19}. 

To provide chemical support for the sibling scenario we, i) explored the chemical similarities between the two GCs, ii) compared each to the disc-like GC NGC~6752 and iii) compared differences between NGC~288 and NGC~362 with the differences seen in the three LMC GCs NGC~1786, NGC~2210 and NGC~2257 as well as the LMC ``blood tied'' GCs, NGC~2136 and NGC~2137. To compare our two GCs, we first split NGC~362 into two groups separated by average $s$-process abundance, following the discovery and designation made in \citetalias{montypaperI}. In \citetalias{montypaperI}, the $s$-weak group was found to be older than the $s$-rich group and to share greater chemical similarities with NGC~288. Therefore, we compared the $s$-weak group and NGC~288 to probe the primordial chemical similarities between the two GCs. 

Overall, we found a maximum disagreement of no more than $\sim0.2$~dex between the $s$-weak group in NGC~362 and NGC~288. The largest disagreement occurred in Nd (an element with both $s$ and $r$ process contributions.) Comparing the $s$-rich group with NGC~288 showed larger disagreement ($>0.2$~dex) with the heavy elements (heavier than Y) but small disagreements otherwise. Both GCs showed large differences in every element when compared to NGC~6752 \citep[using data from the study of][]{yong2013}. When comparing differences between the LMC siblings using data from \citet{mucciarellithreegcs}, to the differences between the $s$-weak group and NGC~288, the $s$-weak group and NGC~288 showed equivalent or greater chemical similarity across nine of the thirteen (9/13) elements in common between studies. However, both groups in NGC~362 showed larger disagreement with NGC~288 in $r$-process abundances compared to those seen in the three LMC siblings. In the case of the comparison between the $s$-rich group in NGC~362 and NGC~288, all of the heavy element abundances showed larger disagreements compared to those seen in the LMC siblings.

Dynamical evidence against the sibling scenario centred primarily around an investigation into the different apparent dynamical states of the two. Specifically, that NGC~362 is a suspected post-core-collapse (PCC) GC while NGC~288 is not \citep{harris96}. Characteristics of LMC GCs from the study of \citet{mackey03} were used to explore the relationships between mass, age, radial distance from the LMC centre and cluster core collapse. PCC GCs in the LMC were among the oldest and most massive GCs, supporting the idea that they are among the most dynamically evolved. However, non-PCC clusters were also observed among the oldest LMC GCs, suggesting that GCs could survive to present day without significant dynamical evolution under the assumption of differing initial conditions (mass, density and stellar populations). Therefore, we suggest that significantly different initial conditions are required to explain the near-identical orbits but vastly different dynamical stars of NGC~288 and NGC~362. This is supported by existing $N$-body simulations of the clusters \citep{baumgardthilker}.

Chemical evidence against the sibling scenario stemmed mainly from further exploring the disagreement in heavy elements discovered in Sec.~\ref{sec:chemsupport}. This involved comparing with the chemical abundances of GSE stars from the studies of \citet{matsuno21} and  \citet{aguado21}, who studied Gaia-Enceladus and Gaia-Sausage stars, respectively. Given that Gaia-Enceladus includes the entirety of Gaia-Sausage stars, we refer to both as ``GSE'', as is common in the literature. GSE stars were found to be enhanced in the $r$-process element Eu and dominated overall by the $r$- compared to the $s$-process. The primordial, $s$-weak population in NGC~362 was found to be more strongly dominated by the $r$-process than NGC~288 (with a maximum difference of $\Delta^{\mathrm{Y/Eu}}=-0.26$, compared to $\Delta^{\mathrm{Ba/Eu}}=-0.10$ in NGC~288). The ratio of $r$-process to $\alpha$ elements was also used to explore an association with GSE, following the discovery by \citet{matsuno21} that GSE stars show an enhancement in [Eu/Mg] ([Eu/Mg]~$\sim0.40$~dex in GSE stars vs. [Eu/Mg]~$\sim0.20$~dex in in-situ stars at [Fe/H]~$\sim-1.3$~dex). We found that both groups in NGC~362 showed a higher ratio of $\Delta^{\mathrm{Eu/Mg}}$ overall, with an average enhancement of $\sim0.1$~dex relative to NGC~288. When considering the better measured $\alpha$-element, Si, the average value of $\Delta^{\mathrm{Eu/Si}}=0.27$ in the $s$-weak group and $\Delta^{\mathrm{Eu/Si}}=0.09$ in NGC~288. 

To confirm these differences, stars from the two clusters were recovered in the GALAH DR3 catalogue \citep{galahdr3}, with the GALAH abundances confirming that NGC~362 was enhanced in both ratios relative to NGC~288. Overall, NGC~362 showed greater agreement with GSE stars when considering trends in the ratios of $s$-, $r$- and $\alpha$ elements. Finally, when comparing the $s$-weak group in NGC~362 directly to NGC~288, the two showed a marked difference in the ratio of first peak $s$-process elements (primarily in Y) to $r$-process elements (Eu), suggesting primoridal differences between the two clusters.

Finally, we experimented with fitting the chemical abundance ratios in the two GCs with a galactic chemical evolution (GCE) model. The model was created under a closed-box assumption using the \textsc{python} chemical evolution package \texttt{OMEGA} \citep{omega}. We adopted as our base model, the GCE model of \citet{matsuno21} and explored the fit to six different element ratios representing the combination of four different nucleosynthetic groups ($\alpha$:~\ion{Si}{I}, Fe-peak:~\ion{Fe}{I}, $s$-:~\ion{Ba}{II} and $r$-process:~\ion{Eu}{II}). A selection of GSE stars were also selected from the study of \citet{buder22} and the GMM predictions from the study of \citet{myeong22} were also used to help evaluate the fit of the GCE model. In almost all element ratios, one or both $s$-process groups in NGC~362 showed excellent agreement with the GALAH GSE star averages. In the case of $s/\alpha$ and $r/\alpha$, the two were overlapping. NGC~288 showed less agreement overall, with the closest agreement being in the ratio of $s/r$ where the two showed a difference of $\sim0.1$~dex.

The base GCE model was found to under-produce in the $r$-process element Eu relative to the $\alpha$ element Si leading to the exploration of three additional models to improve the fit. Overall, the best-fit GCE model to the GALAH GSE datapoints was found to be Model 3 at higher metallicities (NGC~362-like) and Model 4 at lower (NGC~288-like) metallicities. Models 3 and 4 saw the fraction of neutron star mergers increase by 0.25\% and 0.5\% and the turnover mass from AGB to massive stars increase by two and four solar masses respectively from the baseline model. Model 3 was found to fit primordial $s$-weak group in NGC~362 remarkably well in the ratio of [$r$/$\alpha$] and [$s$/$\alpha$]. 

\section{Conclusions}
\label{sec:conc}
In conclusion, although the clusters show remarkable chemical similarities and appear dynamically coincident to the GSE merger in integrals of motion space, they show distinct differences in $r$-process element abundances. The aforementioned chemical ratios provide a clear link between NGC~362 and GSE and Sausage stars, this is not the case for NGC~288. When fitting the two GCs with a GCE model designed to fit GSE stars, NGC~362 shows excellent agreement with a slightly altered model, while NGC~288 does not. We hypothesise that the two are either i) \textit{not} galactic siblings, and were therefore brought in via two separate, but perhaps similar events, or ii) that chemical in
-homogenities across the GSE progenitor were large enough to create distinctions between GCs and galactic stars born at different spatial locations. This second point demands a larger degree of chemical inhomogeneity in the GSE progenitor than what is presently seen in the data. In the future, we would like to build a larger database of high precision comparisons between known GC siblings (found in-situ around their birth galaxy) and/or comparisons between GCs and galactic member stars.

The two papers in this series represent the first high-precision differential chemical abundance analysis of multiple MW GCs. With errors as small as 0.01~dex we have delivered the most precise comparative/relative chemical abundance study ever undertaken of GCs. Coupled with high precision 6D phase space measurements from Gaia, and thus orbits, this study is the most comprehensive relative chemo-dynamical GC study thus far. While it is not possible to unambiguously confirm or refute the conjecture that NGC~288 and NGC~362 are siblings, this pioneering study is a ``proof of concept'' demonstration and builds the conceptual and analysis framework for future studies in this area.

In conclusion, NGC~362 and NGC~288 are remarkably similar chemically, but show distinct differences that dispute their joint association with GSE. We propose that either the two are not siblings, or that GSE was chemically in-homogenous enough to explain the disagreements.

\section*{Acknowledgements}
SM also wishes to acknowledge the traditional custodians of Mt. Stromlo, the Ngunawal and Ngambri people and pay her respect to elders past and present. SM acknowledges funding support from the Natural Sciences and Engineering Research Council of Canada (NSERC), [funding reference number PGSD3 - 545852 - 2020]. Cette recherche a été financée par le Conseil de recherches en sciences naturelles et en génie du Canada (CRSNG), [numéro de référence PGSD3 - 545852 - 2020]. This research was supported by the Australian Research Council Centre of Excellence for All Sky Astrophysics in 3 Dimensions (ASTRO 3D), through project number CE170100013.

This work relies heavily on the \texttt{Astropy} \citep{astropy1, astropy2}, \texttt{SciPy} \citep{scipy}, \texttt{NumPy} \citep{numpy} and \texttt{Matplotlib} \citep{matplotlib} libraries and \texttt{Jupyter} notebooks \citep{jupyter}. Based on observations collected at the European Organisation for Astronomical Research in the Southern Hemisphere under ESO programme 075.D-0209(A)
%%%%%%%%%%%%%%%%%%%%%%%%%%%%%%%%%%%%%%%%%%%%%%%%%%
\section*{Data Availability}
The data underlying this article are available in \citetalias{montypaperI} in the supplementary material provided.



\bibliographystyle{mnras}
\bibliography{mnras_template_updated} % if your bibtex file is called example.bib

%\begin{thebibliography}{99}
%\end{thebibliography}

%%%%%%%%%%%%%%%%%%%%%%%%%%%%%%%%%%%%%%%%%%%%%%%%%%

%%%%%%%%%%%%%%%%% APPENDICES %%%%%%%%%%%%%%%%%%%%%

\appendix

\section{Additional Chemical Correlations}
\label{app:allchemcorr}
\begin{figure*}
\centering
\includegraphics[scale=0.6]{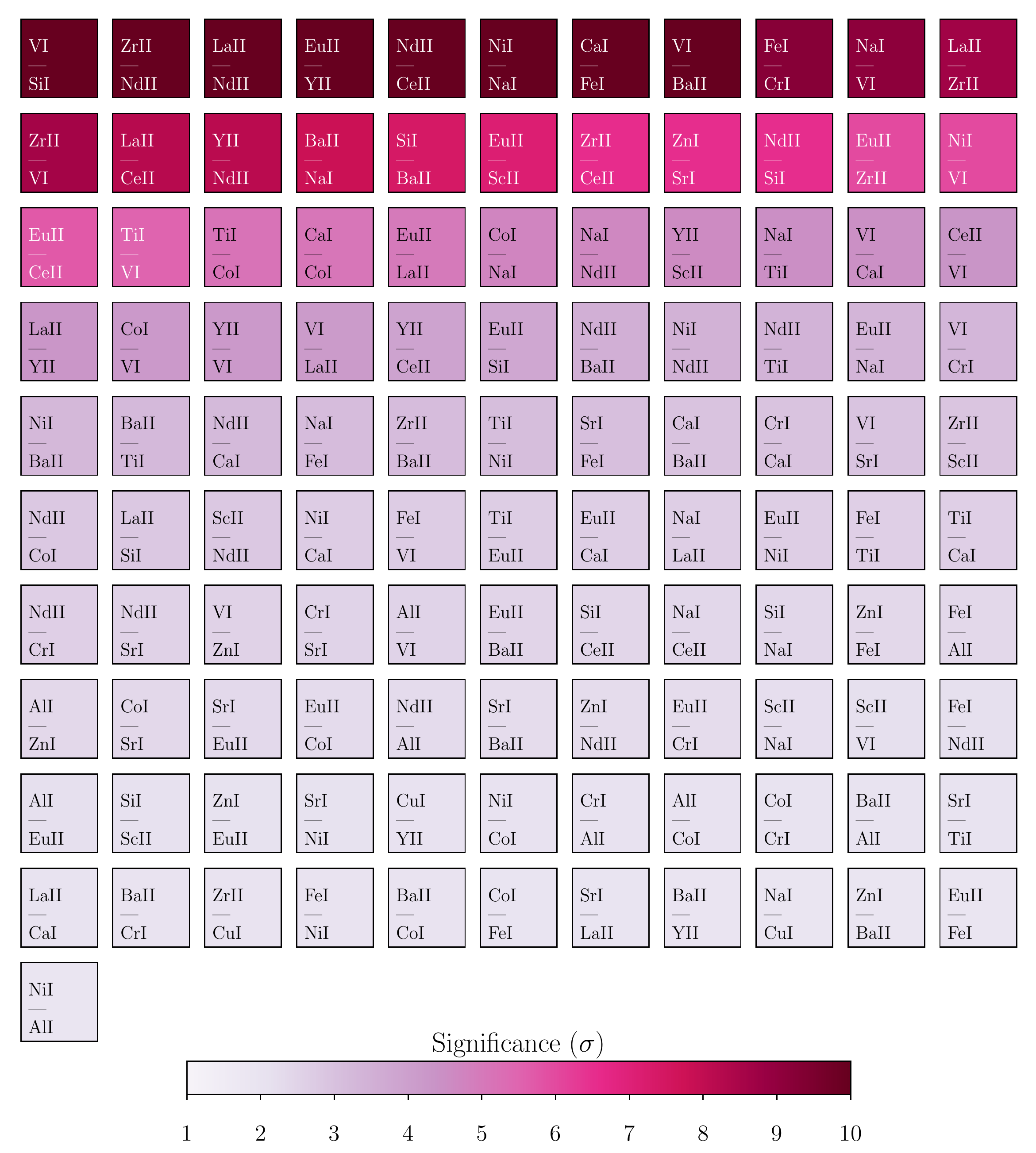}
    \caption{All possible element ratios coloured the statistical significance of the difference between the primordial $s$-weak group in NGC~362 and all stars in NGC~288. Note the differences in the ratio of the first peak $s$-process element Zr and the $r$-process element Nd between the clusters in addition to the differences in Y and Eu discussed in Sec.\ref{sec:chemevidagainst}.}
    \label{fig:allelemratios}
\end{figure*}

Fig.~\ref{fig:allelemratios} presents the differences between the two clusters in all possible element ratios, each coloured by significance. Interpretations for Fig.~\ref{fig:allelemratios} are discussed in Sec.~\ref{sec:chemevidagainst}.

\section{Exploring Orbital Changes to NGC~362 Under the Addition of MW Bar}
\begin{figure*}
	\includegraphics[scale=0.325]{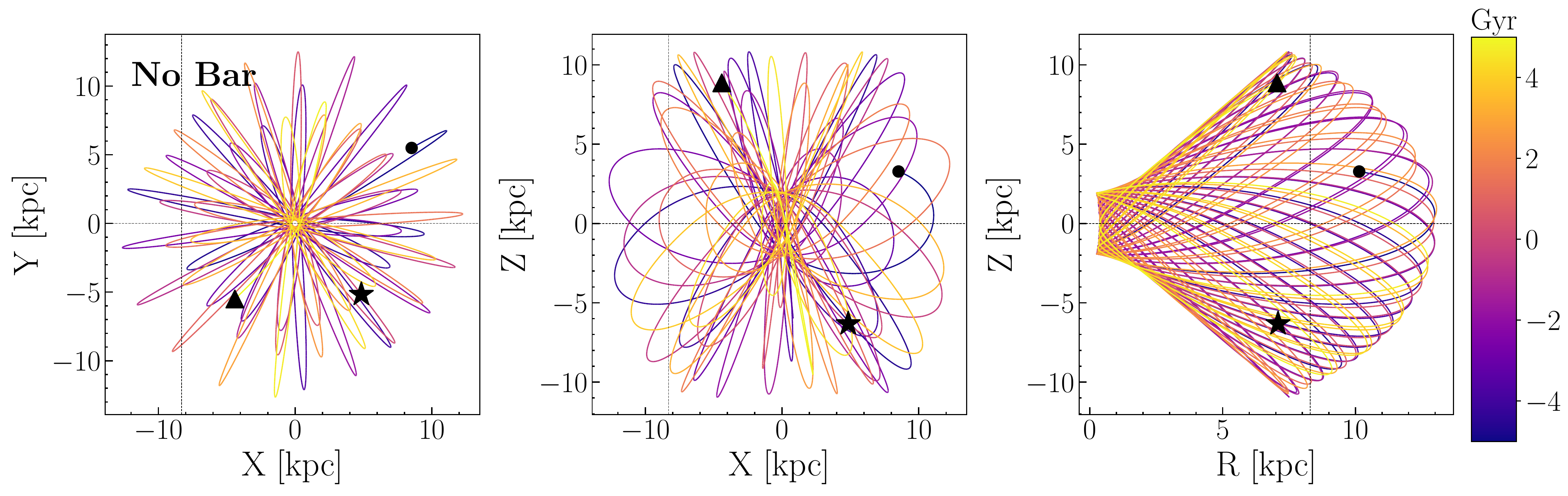}
	\includegraphics[scale=0.325]{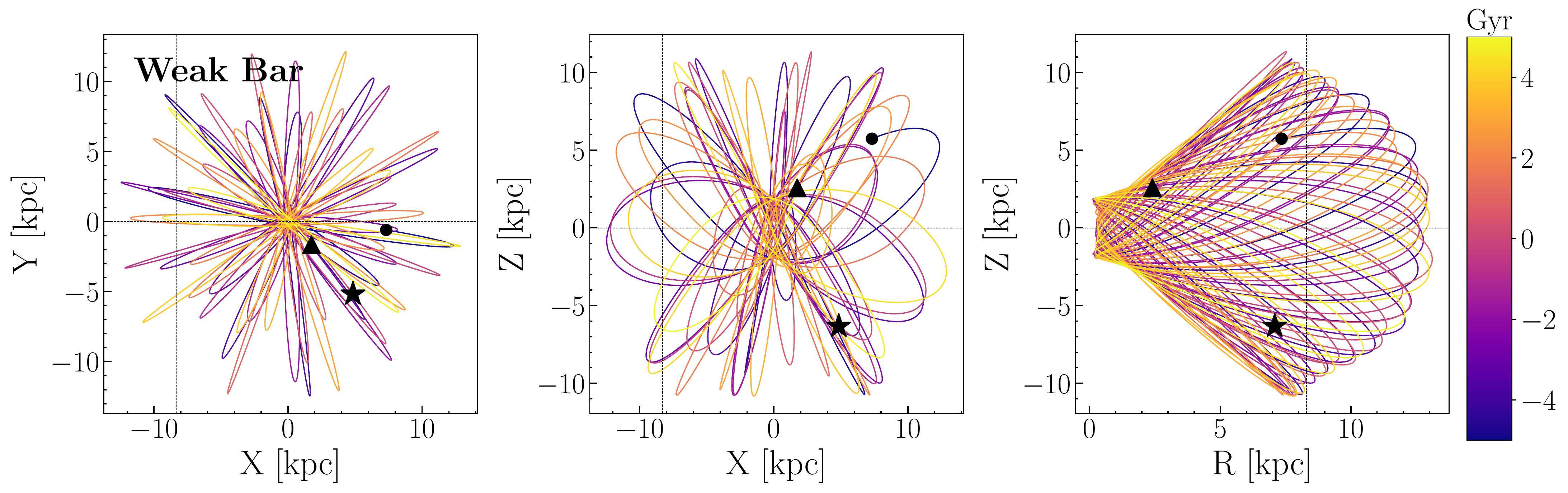}
	\includegraphics[scale=0.3275]{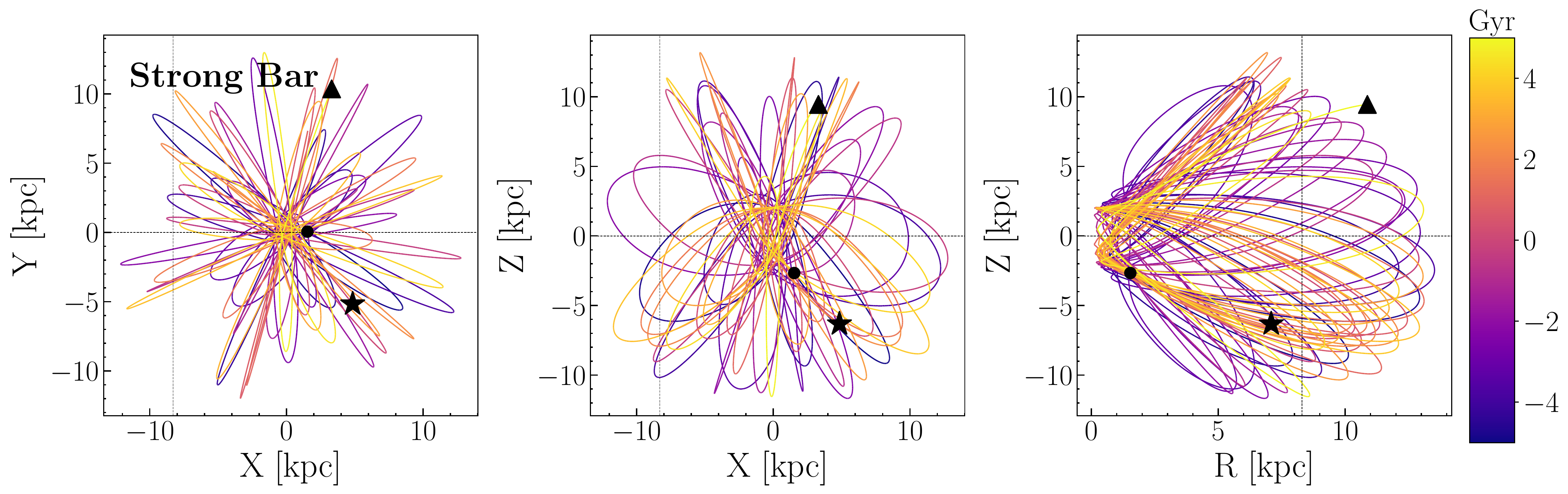}
    \caption{\label{fig:ngc362orbs} As in Fig.~\ref{fig:orbitdifs}, for the GC NGC~362 integrated forwards and backwards for 5~Gyr in three different potentials. The position of the cluster 5~Gyr ago is marked with a circle, the present day position is marked with a star and the position 5~Gyr in the future is marked with a triangle. The trajectory is coloured by time and presented (from left to right) in the Y-X, Z-X and Z-R planes. \textit{Top}, the orbit integrated in \texttt{MWPotential2014}, \textit{middle}, integrated in \texttt{MWPotential2014+DehnenBarPotential} with weak bar characteristics, \textit{bottom} integrated in \texttt{MWPotential2014+DehnenBarPotential} with strong bar characteristics. Details of the bar characteristics are given in Section~\ref{sec:dynsib}. The position of the sun is marked with dashed lines.}
\end{figure*}

Fig.~\ref{fig:ngc362orbs} shows the changes to the orbit of NGC~362 under three different MW potentials.

%%%%%%%%%%%%%%%%%%%%%%%%%%%%%%%%%%%%%%%%%%%%%%%%%%

% Don't change these lines
\bsp	% typesetting comment
\label{lastpage}
\end{document}